\documentclass[traditabstract]{aa}
\usepackage{txfonts}
\usepackage{graphicx}
\usepackage{natbib}
\usepackage{epstopdf}
\usepackage{verbatim}

\begin{document} 

\title{The VISTA Science Archive}
 
\author{N.J.G.~Cross\inst{1} \and R.S.~Collins\inst{1}
 \and R.G.~Mann\inst{1} \and M.A.~Read\inst{1} \and
E.T.W.~Sutorius\inst{1} \and R.P. Blake\inst{1} \and M. Holliman\inst{1} \and N.C.~Hambly\inst{1}  
\and J.P.~Emerson\inst{2} \and A.~Lawrence\inst{1} \and K.T.~Noddle\inst{1}}

\institute{Wide-Field Astronomy Unit, Institute for
Astronomy, School of Physics and Astronomy, University of Edinburgh, Royal Observatory, Blackford Hill, Edinburgh EH9 3HJ
\and Astronomy Unit, School of Physics \& Astronomy,
     Queen Mary University of London, Mile End Road, London E1 4NS}

\date{Accepted XXXX. Received XXXX ; in original form XXX }


\abstract{
We describe the VISTA Science Archive (VSA) and its first public release of data
from five of the six VISTA Public Surveys. The VSA exists to support the VISTA 
Surveys through their lifecycle: the VISTA Public Survey consortia can use it
during their quality control assessment of survey data products before submission to 
the ESO Science Archive Facility (ESO SAF); it supports their exploitation of
survey data prior to its publication through the ESO SAF; and, subsequently, it 
provides the wider community with survey science exploitation tools that
complement the data product repository functionality of the ESO SAF. 

This paper has been written in conjunction with the first public release of 
public survey data through the VSA and is designed to help its users understand
the data products available and how the functionality of the VSA supports their 
varied science goals. We describe the design of the database and outline the 
database-driven curation processes that take data from nightly 
pipeline-processed and calibrated FITS files to create science-ready survey 
datasets. Much of this design, and the codebase implementing it, derives from
our earlier WFCAM Science Archive (WSA), so this paper concentrates on the
VISTA-specific aspects and on improvements made to the system in the light of
experience gained in operating the WSA.} 


\maketitle

\label{firstpage}


\section{Introduction}  \label{intro}
One of the clearest trends in observational astronomy over the past two decades
has been the rise in the importance of systematic sky surveys. When coupled with
good archives, sky surveys provide homogeneous, re-usable data products 
facilitating a range of research programmes and, in particular, enabling the 
large-scale statistical analyses required for many of the most important science
goals in modern astronomy.

Probably the most prominent of this generation of surveys has been the Sloan 
Digital Sky Survey \citep[SDSS;][]{SDSS} and its SkyServer archive system 
\citep{SkyServer} demonstrated the power of a survey archive based on a
Relational Data Base Management System (RDBMS). Such systems can offer
astronomers the ability to pose powerful analytical queries in Structured Query
Language (SQL) against seamless, survey-wide source catalogues, thereby 
enabling survey science that would be impossibly cumbersome for an astronomer 
provided with nothing more than a repository of matching image and catalogue 
files for each of the thousands of pointings making up the survey dataset.

The success of the SDSS SkyServer was a strong influence on the design of the 
archive component of the VISTA Data Flow System \citep[VDFS;][]{VDFS}. VISTA,
the Visible and Infrared Survey Telescope for Astronomy, is described by
\cite{VISTA}. VDFS was designed as a two-phase project, with an initial goal of
supporting near-infrared surveys to be conducted with the Wide Field CAMera 
\citep[WFCAM;][]{WFCAM} on the UK Infrared Telescope (UKIRT) and an ultimate 
objective of supporting surveys with VISTA. Within VDFS, the Cambridge Astronomy
Survey Unit (CASU) run a night-by-night data processing pipeline, with the 
Wide-Field Astronomy Unit (WFAU) in Edinburgh generating further data products 
and providing science archive facilities.

The first-generation VDFS archive -- the WFCAM Science Archive (WSA) -- is 
described by \cite{WSA} and serves catalogue and image data from the UK Infrared
Deep Sky Survey \citep[UKIDSS;][]{UKIDSS}, as well as other, P.I.-mode, data
taken with WFCAM. More than 1000 users are registered for authenticated access to 
proprietary UKIDSS data through the WSA, and it supports anonymous access by a 
larger community once the data are public. The phased approach adopted within 
the VDFS ensured that the design and development of the WSA progressed with 
scalability to the larger data volumes of VISTA kept explicitly in mind, along 
with the likely scientific usage patterns of the VISTA surveys. For example, 
\cite{Crs09} describe the enhancements made to the WSA database schema to 
support time-series analysis of multi-epoch data, which was prototyped using 
observations from the UKIDSS Deep Extragalactic Survey, but motivated by the 
requirements for supporting variability analyses with VISTA.

The initial scientific programme for VISTA is mostly focussed on six ESO Public 
Surveys which deliver reduced images and derived catalogue data products to the
ESO Science Archive Facility (SAF). Five of the six Public Survey consortia 
(see \S~\ref{sec:vista}), use the VDFS for the generation of these data products
and employ the VISTA Science Archive (VSA) to manage their data, both for 
quality assurance analysis and preliminary exploitation prior to submission to
the ESO SAF and, following its publication there, to provide the wider community
with sophisticated science archive capabilities that complement the data product
repository functionality of the ESO SAF.

\subsection{Summary of earlier VDFS work}

Since the design of the VSA is derived so directly from that of the well-used
WSA, we strongly urge readers who are unfamiliar with the VDFS science archives
to read the existing papers on the WSA \citep{WSA,Crs09}, as well as the
comprehensive online documentation provided on the VSA 
website\footnote{http://surveys.roe.ac.uk/vsa}, in conjunction with this paper.
However, we will summarise the main points of these papers which are relevant
to the VSA in this section.

VISTA and UKIRT/WFCAM take very similar basic data, which consists of images
obtained by a wide-field near infrared imaging array on a $\sim$4m telescope. 
The two cameras utilise large format 2048$\times$2048 pixel infrared
detectors: sixteen 0.34 arcsec pixel$^{-1}$ in the case of VISTA, and four 0.4
arcsec pixel$^{-1}$ in the case of WFCAM. Images are taken in 5
broad-band filters ($Z$, $Y$, $J$, $H$, $K_s$/$K$) and several narrow-band
filters, within the wavelength range $0.8<\lambda<2.5\mu$m, for a range of large
surveys and smaller P.I. programmes. Large areas can be covered by 
repeating a basic pattern of images, a square pattern of 4 adjacent pawprints 
in the case of WFCAM.
 
Observing time is divided between large surveys (UKIDSS and the Campaigns on
UKIRT/WFCAM) and smaller PI-led programmes (time awarded by a telescope
allocation committee), service mode observations for very small projects and 
special projects like director's discretionary time projects and calibration 
work. WFAU ingest all of these data into the WSA but the key design 
features were driven by the requirements for the main surveys: the UKIDSS
surveys and the then future VISTA Public Surveys.

In VDFS, a single exposure, reduced science frame is designated as a {\it
normal} frame. These frames can be stacked (coadded to increase the
signal-to-noise) together with small offsets in position (a jitter pattern) to
reduce the effects of bad pixels: the resulting image frame is designated a 
{\it stack} frame. The pixel size of WFCAM images is $0.4\arcsec$, but the 
seeing sometimes is as good as $0.5\arcsec$, so the images can be undersampled. 
To produce critically sampled images in the best seeing conditions, a technique
called micro-stepping was used, where a series of images are interleaved; 
adjacent frames that are offset by half or a third of a pixel, (in reality a 
small integer offset is added too to reduce the affects of bad pixels) in the x
and y directions, to create a $2\times2$ ({\it leav}) image with $0.2\arcsec$
(or $3\times3$ with $0.133\arcsec$) pixels. Several {\it leav} frames with 
different jitter positions are stacked together, the resulting image frame is 
designated a {\it leavstack} frame.
 
A {\it stack} ({\it leav} or otherwise) is the fundamental science image, from
which catalogues are extracted. If there are multiple epochs in the same
filter and pointing, a {\it deepstack} or {\it deepleav\-stack} may be created
to find fainter sources. All of these science frames have associated confidence
({\it conf}) images that contain the relative weighting of each pixel. This
includes bad pixel masking, the effects of jittering, seeing and exposure time 
weighting. 

At WFAU, for the purposes of archiving, UKIDSS as a whole was referred to as a
{\it survey} while the different parts of it: the Large Angular Survey (LAS),
the Galactic Plane Survey (GPS), the Deep Extragalactic Survey (DXS) etc. were
referred to as {\it programmes} and PI-led programmes were referred to as 
non-survey programmes. Since all UKIDSS programmes were released together and 
there was a management structure covering the whole of UKIDSS, the distinction 
between {\it survey} and {\it programme} was useful and clear. UKIRT divide 
observations into {\it projects} which have their own identifiers. We assign 
{\it projects} to {\it programmes} based on the metadata identifiers, e.g. in 
the header, the {\bf PROJECT} keyword u/ukidss/gcs5 is UKIDSS GCS project 
observing set number 5 and is assigned to programmeID=103 in the WSA.

The data in the WSA are stored in a relational database management
system (RDBMS), which stores data in a set of related tables. In an RDBMS,
missing values can be handled in a couple of ways: by setting them as NULL, or 
giving them a default value. Setting them as NULL has the advantage that it is 
clear that data is missing, but has the disadvantage that all queries have to 
consider NULL values. For example, if you are querying for objects with a 
certain colour range, $-1.0<$(J-H)$<1.0$, it would be necessary to include a 
constraint ``and (J-H) is not null'', which would occur if one or other of the
two images had not yet been observed or the object was too faint in the $J$ or
$H$ band for a detection. Our solution is to use default values, but our
standard default values are large negative values, that are outside the normal 
range of most physical values. The default values for each column are given in
the schema browser, see below. All of the columns in all of the tables are {\it
not null}, i.e. they have at least a default value.

The pixel data from images is not stored directly in the RDBMS as binary large
objects (BLOBs), but they are stored as flat files in multi-extension FITS 
\citep[MEF; a FITS file,][with a primary which contains metadata about the
whole file, and secondary extensions with binary data from each detector and a header
with detector relevent attributes]{FITS} format, and the paths are stored in the
archive, along with all of the metadata. The databases are self describing, 
since all the information about the surveys is also stored in the database, 
in curation tables, which are database tables that are designed principally to 
aid the curation of each programme, see \S~\ref{sec:curTab}.

The design of the relational database should capture the inherent structure of
the data stored. For instance the image metadata can be divided into different
groupings, e.g. metadata related to the whole MEF, such as observation time,
filter, project, PI, airmass and metadata related to each detector extension,
such as sky level, zeropoint, seeing. When designing a relational database, we capture
the structure in an entity-relation model (ERM), see Fig.~\ref{fig:tilepawERM}
for an example. An ERM contains {\it entities} (shown as boxes with rounded
corners) which represent a collection of related data, e.g. primary header data from each FITS image or sources in a merged
filter catalogue. The relationship between {\it entities} is represented by
lines between them which are mandatory (solid line), optional (dotted) and can be one-to-one (a single line), one-to-many (a single
line that branches into three, similar to a crow's foot) or many-to-many (three
lines converge to a single line that branches into three). If the two tables
share the same primary key a perpendicular line is added across the 
main line. The basic ERMs for the WSA (which are relevant for the VSA) are shown in
\cite{WSA} and we show the multi-epoch ERMs in \cite{Crs09}.

While the database models presented so far could be implemented in any
RDBMS, the WSA and VSA are implemented in a commercial software product, 
Microsoft SQL Server, which is suitable for medium to large scale databases.
This was also the choice that the Sloan Digital Sky Survey team made, which
heavily influenced our decision. When we implement the data model the ERMs are
converted into a schema, a set of database objects. Most of these objects are
tables, with entities in the ERM mapping to tables in the schema. The
tables hold all the data and can be queried via the user interface applications. 

The WSA provides a schema browser, which contains all the information about all
of the tables in each release. The left hand side gives the user a list of 
surveys and releases and then the tables, functions and views associated with 
each one. Selecting a table gives a description of the table and then the list 
of attributes in the table. Each attribute has the name, type (e.g. int, real, 
float), length (in bytes), units, description, default value and unified content
descriptor. Each table has a primary key, which is a single attribute or a
combination of attributes in the table that between them can cover all unique
entries and must be unique. For instance, the \verb+Multiframe+ table contains
entries for each multi-extension FITS image, so the primary key is the
{\bf multiframeID}, whereas in the \verb+lasDetection+ table, entries are
objects extracted from each extension of a FITS image, and within each
extraction are given a sequence number, so the primary key is ({\bf
multiframeID}, {\bf extNum}, {\bf seqNum}). Any constraints with 
respect to other tables are listed at the top and all attributes which are 
indexed for fast searches are highlighted. The primary key attributes are 
indexed automatically. Some attributes require a more detailed description, and 
these are linked to a glossary and have a symbol like a book beside the 
attribute name, which can be clicked on. 

Other database objects include Views and Functions. Views are selections of data
from tables already in the database. They can be a subset from one table or a
superset of many, and are queried in the same way as tables, but no extra data
is stored. These are used, for instance, in the WSA when we give a subset of the
source tables for some UKIDSS surveys which have data taken in all filters. 
Functions take inputs and do specific calculations. We have some which do 
spherical astronomy calculations, give expected magnitude limits, convert ra 
and dec to a sexagesimal string and give names for objects in the IAU 
convention.

\subsection{Outline of paper} 

This paper describes the VSA and its first public release of data from the five 
VDFS-supported Public Surveys. In Section~\ref{sec:vista} we discuss the
VISTA telescope and the Public Surveys and compare to UKIRT-WFCAM, focussing
on the essential differences that affect the VSA. In Section~\ref{sec:overview},
we provide an overview of the VSA, discussing the table structure before we
compare the VSA to the WSA in Section~\ref{sec:differences}. We discuss changes 
to the image metadata, the catalogue parameters and the infrastructure in
Sections~\ref{sec:im_meta_ch}--\ref{sec:infra} in the VSA compared to the WSA
and new features that are common to both in Section~\ref{sec:newFeatures}.
Section~\ref{sec:queries} provides examples of some of the different types of 
science queries that the VSA supports, while Section~\ref{sec:releases} provides 
details of the contents of the first VSA releases of the five VDFS-supported 
VISTA Public Surveys. We summarise this paper and discuss future work in 
Section~\ref{sec:summary}, while several appendices provide technical details 
supplementing the main body of the paper. 

\section{Overview of VISTA and its Public Surveys}
\label{sec:vista}

The Visible and Infrared Survey Telescope for Astronomy \citep[VISTA; ][]{VISTA}
is currently the fastest near-infrared survey telescope, with an \'{e}tendue
(area times instantaneous field-of-view) of approximately $6.5$m$^2$deg$^2$. It
has a 4m f/1 primary mirror, and a 1.2m secondary giving it a 1.65 degree diameter field-of-view \citep[see][]{EmerSuth2010a,EmerSuth2010b}. The VISTA Infra Red CAMera
\citep[VIRCAM;][]{Dalt10}, has 16 $2048\times2048$ pixel non-buttable Raytheon
VIRGO HgCdTe detectors and has a quantum efficiency $>80\%$ between $0.9\mu$m 
and $2.4\mu$m. The pixel scale is $0.34\arcsec$ and the instantaneously sampled 
field-of-view is 0.6 sq. deg (see Fig.~\ref{fig:VISTAfp}). Compared to its
nearest counterpart, the United Kingdom Infra Red Telescope with its Wide Field CAMera 
\citep[UKIRT-WFCAM;][]{WFCAM}, the survey speed of VISTA is $\sim6$ times
faster, having twice the sensitivity -- increased throughput for a similar 
sized telescope -- and 3 times the area per pointing.

ESO's Science Verification for VISTA started at the end of 2009 
and the main science programme commenced in early 2010. VISTA's programme 
initially focuses on six ESO Public Surveys\footnote{http://www.eso.org/public/teles-instr/surveytelescopes/vista/
surveys.html}, nicely complementing the sub-surveys of UKIDSS in the northern 
sky.

These six surveys are:
\begin{itemize}

\item VHS: the VISTA Hemisphere Survey\footnote{http://www.ast.cam.ac.uk/$\sim$rgm/vhs/};

\item VVV: the VISTA Variables in Via Lactea\footnote{http://vvvsurvey.org}
\citep{VVVDR1};

\item VMC: the VISTA Magellanic Cloud 
survey\footnote{http://star.herts.ac.uk/$\sim$mcioni/vmc/} \citep{VMC};

\item VIKING: the VISTA Kilo-degree INfrared survey for
Galaxies \citep{VIKING};

\item VIDEO: the VISTA Deep Extragalactic Objects
survey\footnote{http://star-www.herts.ac.uk/$\sim$mjarvis/video/} \citep{VIDEO};

\item UltraVISTA\footnote{http://www.strw.leidenuniv.nl/$\sim$ultravista/}
\citep{ULTRAVISTA}.
  
\end{itemize}

These surveys have a `wedding cake' arrangement of galactic/extragalactic
surveys (VHS, VIKING, VIDEO, UltraVISTA) with different depth/area combinations 
and two specialised stellar astronomy programmes (VVV, VMC), much
like the UKIDSS surveys. The five surveys supported by the VDFS are VHS, VVV, VMC, VIKING
and VIDEO. UltraVISTA makes use of the CASU pipeline products, but is not
currently archiving its data in the VSA.

VISTA data is calibrated on the natural VISTA photometric system (see Hodgkin
et al. 2012, in preparation). All magnitudes (unless designated as AB mag) are
on this Vega mag system. 

\subsection{Differences in the telescope and instrument between VISTA \& WFCAM}

For detailed descriptions of VISTA and WFCAM, see \cite{VISTA} and \cite{WFCAM}
respectively. In this section we will just discuss the salient differences which
affect the VSA design compared to the WSA. 

\subsubsection{The VISTA focal plane: pawprints and tiles}
\label{sec:tile_pawprint}

VISTA is significantly different from UKIRT/WFCAM in several important aspects,
which affect image processing and subsequent archive operations. The
most significant differences of VISTA to UKIRT/WFCAM are the arrangement
of the focal plane and the ability VISTA's alt-azimuth mount provides to observe the same piece of sky in any orientation with respect to the focal plane.

\begin{figure}
\resizebox{\hsize}{!}{\includegraphics{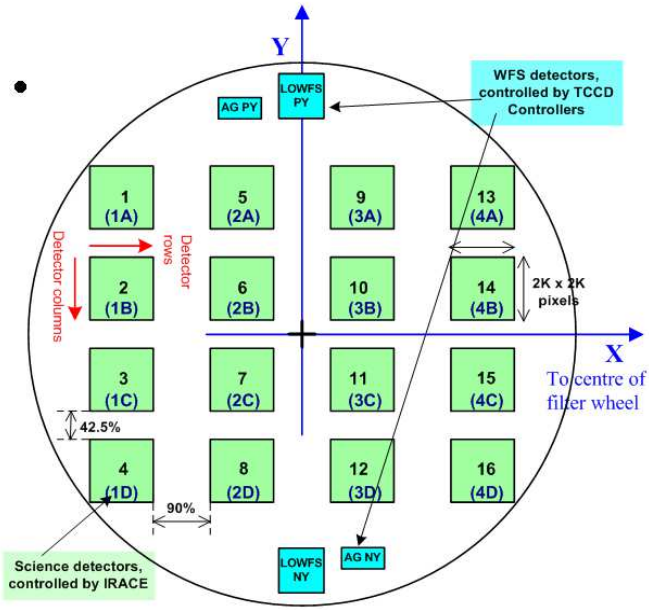}}
  \caption{The VISTA focal plane showing 16 2k $\times$ 2k detectors with 90\%
  spacing in the x-direction and 42.5\% in the y-direction. There are also two
  auto-guider (AG) and two low-order wave-front sensor (LOWFS) detectors.}
\label{fig:VISTAfp}
\end{figure}

\begin{figure}
\resizebox{\hsize}{!}{\includegraphics{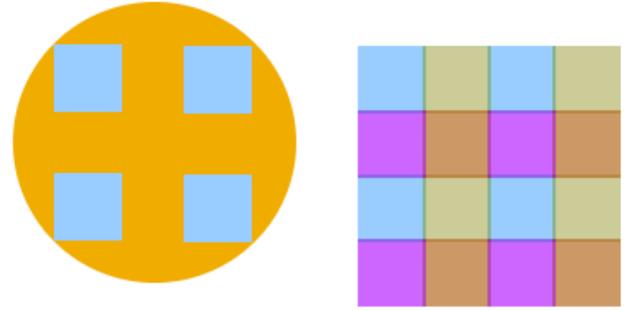}}
  \caption{Areas of sky can be efficiently surveyed by arranging 4 WFCAM
  pawprints (left) into the arrangement on the right. Each colour in the right
  hand image represents a different pawprint. There is a small amount of overlap
  at each edge.}
\label{fig:WFCAMfp}
\end{figure}

VISTA has 16 2k$\times$2k Raytheon VIRGO
detectors arranged in a {\em pawprint\/} with detectors spaced 90\%
(10.4\arcmin) of a detector apart in the X-direction and 42.5\% (4.9\arcmin)
apart in the Y-direction (see Fig.~\ref{fig:VISTAfp}) whereas WFCAM has 4 
2k$\times$2k Hawaii 2 detectors arranged in a {\em pawprint\/} (see 
Fig.~\ref{fig:WFCAMfp}) with the same spacing of 94\% (12.8\arcmin) in each
direction.

\begin{figure}
\resizebox{\hsize}{!}{\includegraphics{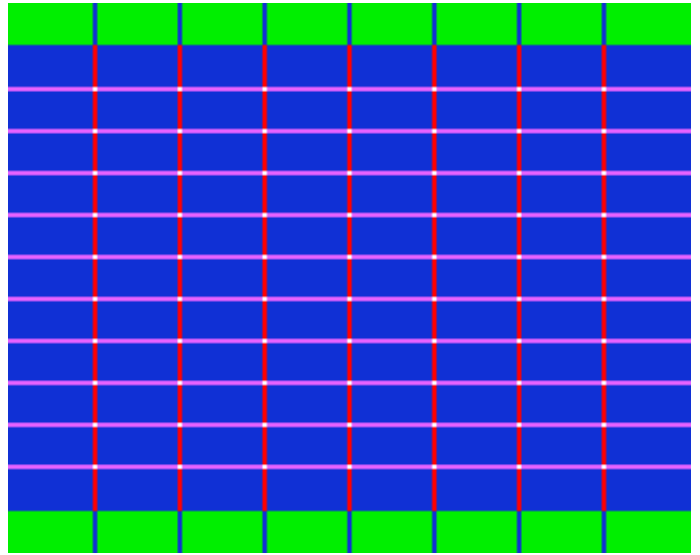}}
  \caption{ An exposure map of a VISTA tile. The green {\it strips} at top \& 
  bottom have a single exposure. The majority of the area (blue) has two  
  exposures, the pink has 3 exposures, the red 4 and the white 6. The doubly
  or more exposed area is 1.501~sq.deg. The singly exposed green {\it 
  strips} at top \& bottom of the plot are each 1.475
  deg$\times$0.092~deg=0.135~sq.deg and can be overlapped by corresponding
  areas from adjacent tiles for many surveys. Assigning one of the two 0.092 deg 
  overlap (top \& bottom) to each of the adjacent tiles involved in an overlap, 
  means that each tile, when part of a filled larger area, covers 
  (1.017+0.092)$\times$1.475=1.636~sq.deg at least twice.}
\label{fig:VISTAtl}
\end{figure}

The VISTA basic filled survey area is a {\it tile} made up of six pawprints, three
in the Y-direction separated by 0.475 of a detector, and two in the X-direction
separated by 0.95 of a detector. Except at two {\it strips} with just a single exposure, this
tile has between twice and six times the exposure time of each pawprint at every
pixel with a mode of two exposures, (see Fig.~\ref{fig:VISTAtl}). There are two
possible ways to achieve a required uniform minimum (two pawprints contributing)
depth across a multi-tile survey. The most efficient in terms of observing time
is for successive tiles to overlap the {\it strips} at top and bottom, coadding
the data from the two separate tiles, and each will have an area of 1.636~sq.deg
covered at least twice. However reaching the full depth by coadding these two 
{\it strips} can be complicated by varying sky conditions if the adjacent tiles
are not observed under the same conditions (e.g. different PSFs and sky
conditions). Indeed the same two effects can be a problem in making a tile from
six pawprints (depending on how rapidly the seeing varies between pawprints).
The other less efficient, but simpler, way to achieve constant depth across a
multi-tile survey is to butt together regions of tiles that have reached the 
minimum double exposure, ignoring the singly exposed {\it strips} resulting in
an area of 1.501~sq.deg covered at least twice.

In the case of WFCAM four pawprints are required to make a filled tile, and
everything gets a single exposure and there are no edge {\it strips}, so large 
contiguous regions of the sky can be surveyed by simply overlapping subsequent
pawprints as shown in Fig.~\ref{fig:WFCAMfp}. We note that in common with WFCAM,
`microstepping' is possible with VISTA, but it is not usually necessary since
the pixel size on VISTA is smaller than WFCAM and the typical seeing at Cerro
Paranal is slightly worse than Mauna Kea, so most images are critically sampled.
Moreover, its use is not recommended by ESO and the VISTA Public Surveys do not
use the technique.

\subsubsection{Merging VISTA pawprints into tiles}
\label{sec:proc_tiles}

The processing of raw pawprint frames into calibrated images is done by the
VDFS pipeline at CASU on a nightly
basis\footnote{http://casu.ast.cam.ac.uk/surveys-projects/vista/technical} and
includes combining the pawprints into tiles. These pawprints and tiles are
ingested into the VSA without any additional image processing. If any pointing 
is observed only once in a given filter in the survey design, the nightly pipeline--processed 
catalogue product is used to generate the merged \verb+Source+ (see
\S~\ref{sec:vsaBasics}) table. For deeper surveys, however, stacking tiles
observed multiple times typically involves observations over several different nights, and multi-night products are not the responsibility of CASU.
WFAU creates tiles from multiple nights of data by first stacking the separate 
component pawprints and then combining the six stacked pawprints into a tile. 
Stacking the tiles rather than the component pawprints is very firmly not
recommended.

While tiling pawprints together to form a tile is quite standard when working
with visible images \citep{MOSAIC}, ground-based infrared sky subtraction is 
difficult because the sky is so much brighter than in the optical and dominates 
the flux of most objects. Furthermore, the VISTA camera's distortion across the 
field-of-view and the larger variation in both the sky and the point-spread 
function (PSF) within the duration of the observations needed to create a 
single tile make it necessary to do additional processing \citep{Lewis10}. 
Tiles are processed using the following procedures\footnote{
http://casu.ast.cam.ac.uk/surveys-projects/vista/technical/tiles gives more
details of the algorithms}:

\begin{itemize}
  \item Stack all components of each pawprint to reach the intended
  signal-to-noise while removing bad pixels by taking the clipped median flux
  value. Each component is shifted by a different number of pixels in the X and
  Y directions as defined by the jitter pattern so that bad pixels or columns
  appear in different positions in each component; 
  \item Extract the catalogue from each pawprint. This detects collections (at
  least 4) of pixels which are all brighter than the background by more that 1.5
  times the sky noise. Overlapping objects are deblended;
  \item Recalculate the world coordinate system (WCS) of each pawprint, by
  comparing stars in the catalogue to the Two Micron All Sky Survey (2MASS)
  catalogue \citep{2MASS};
  \item Calculate the photometric zero-points (VISTA system) of each detector in
  each pawprint and update the headers. For observing block processing this is
  done compared to 2MASS, using the colour equations calculated by CASU and
  whichever stars of the correct colours are in common with 2MASS and the VISTA
  image in question. For deep stacks and tiles, individual stars are compared 
  to the component frames which have already been calibrated using 2MASS. The
  stars in the components and the deep stack are all on the same VISTA system,
  so a direct comparison is possible;
  \item Filter each pawprint to smooth out large scale variations ($>30\arcsec$
  in the background), see \citep{nebul} for details;
  \item Mosaic the 6 unfiltered pawprints into a tile, to produce a tile with
  large scale features present. Mosaicing adjusts all 96 components (16 detectors in
  each pawprint) to the same level and drizzles them on to a single
  tangent-plane projection image;
  \item Mosaic all the filtered pawprints to produce a tile without large-scale
  background features;
  \item Extract the catalogue from the filtered tile;
  \item Recalculate the WCS of the tile; 
  \item Calculate the photometric zero-point (VISTA system) of the tile
  catalogue;
  \item `Grout' the tile catalogue to find the correct PSF in each region of
  the tile and the correct offset for the Modified Julian Date. The grouting 
  procedure tracks the variable flux within the first seven pre-defined
  circular apertures, with radii of $0.5\arcsec$, $\frac{1}{\sqrt{2}}\arcsec$,
  $1\arcsec$, $\sqrt{2}\arcsec$, $2\arcsec$, $2\sqrt{2}\arcsec$ and $4\arcsec$.
  Differential aperture corrections (i.e. the difference between the flux in
  the aperture and the total flux for a point source) are calculated for
  each detector in each pawprint that composes the tile. The fluxes in these 7 
  apertures are then recalculated, although larger aperture fluxes and other
  fluxes, such as the Petrosian flux are not corrected. These larger apertures
  will be only marginally affected by the seeing variations;
  \item Reclassify the stars and galaxies in the tile catalogue. Classification
  originally occurs as part of the extraction process and uses the curve of
  growth to calculate a stellarness-of-profile statistic and classification
  (galaxy 1, star -1, noise 0, saturated -9 , probable star -2 , or probable
  galaxy -3). The different PSFs in each pawprint when combined can give
  misleading classifications, so the classification code takes the grouting
  information about each pawprint and re-estimates the stellarness-of-profile
  statistic and classification;
  \item Recalculate the photometric zero-point (VISTA system) of the tile
  catalogue;
  \item Remove any temporary files, such as the filtered pawprints. 
\end{itemize}

Unsurprisingly the filtering also makes it impossible to produce accurate
catalogue values for large extended sources, e.g.~nearby galaxies or Galactic
nebulae. The filtering scale used for the main VISTA Public Surveys removes
structure on scales larger than $30\arcsec$. This is a similar scale to the
local background scale length ($22\arcsec$) which would in any case limit the accuracy of
any photometry of larger objects. 

\subsubsection{Active Optics}

The quality of VISTA images is maintained, as it observes at different
elevations and its temperature changes, by updating the position of the
secondary mirror, with corrections derived from look up tables and the low order
wave front sensors. If a correction has not been applied recently enough, or it
is bad for some reason, the image quality can be degraded. Some images show
evidence for such problems, contributing, along with seeing variations between
pawprints, and PSF variation across the field when in perfect alignment, to a
greater variety of PSFs than in the case of WFCAM. 

\subsubsection{Telescope mounts}
VISTA is an alt-azimuth mounted telescope, whereas UKIRT has an 
equatorial mount, so the focal plane in WFCAM remains in the same
equatorial orientation, but the VISTA focal plane must rotate with respect to
the telescope to keep the same orientation on the sky during an exposure. Since
the focal plane can rotate, orientation is an additional degree of freedom.
Different programmes can choose the orientation that best accommodates their
survey design. Given the complex processing of image and catalogue data (see
previous subsection), stacking two images at very different orientations was
considered inadvisable. Thus we group data based on orientation as well as 
position/filter when stacking 
and tiling and to do this we have added an extra column, {\bf posAngle} to the 
\verb+RequiredStack+ (see \S~\ref{sec:vsaBasics}) table. This is the orientation
of the image x-axis to the N-S line. This means that if images in the same programme lie in the same 
position and filter but have very different orientations they will be processed 
separately. As a default, we have been using a tolerance of 15 deg, but it can
be set programme by programme. In practice this situation has only occurred once
in the Public Surveys: the VVV team had a small amount of data in the Science
Verification stage which was orientated along equatorial RA axis, and later 
data in the same region of sky which was orientated along the Galactic 
longitude axis, and so the tiles are orientated at $60\deg$ to each other.

\section{Overview of the VSA}
\label{sec:overview}

\subsection{The basic structure of the VSA}
\label{sec:vsaBasics}

WFAU receive VISTA data as multi-extension FITS \citep[MEF; ][]{FITS} files: 
primarily these come from the CASU pipeline, which processes the data by 
Observing Block (OB), but WFAU can also handle `external' FITS images made
outside VDFS, such as those produced by the survey consortia, as is currently 
done with deep VIDEO mosaics, with WFAU merely modifying the headers and
file names prior to ingestion so that the ingestion code can correctly handle
them. The MEFs contain either images or catalogues of objects extracted from
them. Each MEF consists of a primary header, containing metadata relevant to 
the whole observation of the associated pawprint, and one extension for each 
detector of the pawprint, containing metadata relevant to this detector as well
as the binary data (image or catalogue) associated with this detector. In VSA 
parlance the content of an MEF image file is a {\em multiframe\/} (consisting 
of images - frames - for all detectors) and a {\em detection\/} is an object 
extracted from a single image in a single filter. The metadata headers from 
these files are loaded into tables within an RDBMS, as are the catalogue 
tabular data. The header data consists of multiple cards, each no longer than 
80 characters, which contains the keywords with their values and descriptions. 
A standard keyword has a length of 8 characters, but longer, more descriptive 
ones, or even keywords consisting of multiple words can be used by preceding 
them with 'HIERARCH'. HIERARCH keywords originating at ESO start with 'HIERARCH
ESO'.

Depending on the survey, further data products may be generated in a 
database-driven manner. For example, repeated observations of the same field in
the same filter may be stacked to create deeper images, from which catalogues 
are then extracted, while information from multiple different single-filter
catalogues are merged to create catalogues of {\em sources\/}, which are objects
described by attributes in several filters. The metadata from the additional image data products are 
ingested into database tables, as are derived catalogues, while the images 
themselves are stored in FITS format on disk. Further information may then be 
derived from the database tables and stored in new tables: e.g. variability 
information may be derived from multi-epoch data, following the 
synoptic data model of \cite{Crs09}. The VSA comprises, therefore, a set of 
tables within an RDBMS, a collection of FITS files stored on disk and the
interfaces that allow users to access these data.

Whole image files may be selected for download from the archive through web
forms or via SQL queries on the image metadata tables, while image cut-outs may 
be created from these using a different web form. Other web forms exist to 
provide a basic level of access to the catalogue data, but the real power of 
the VSA comes from the ability to query RDBMS tables using SQL. To do this 
requires knowledge of the VSA database schema. The online VSA schema 
browser\footnote{http://surveys.roe.ac.uk/vsa/www/vsa$\_$browser.html} provides 
detailed descriptions of every column in the hundreds of database tables in the 
VSA, but we summarise the five main table classes in the remainder of this 
section.

Throughout this paper, we use a fixed-width font to refer to VSA 
database tables: e.g. \verb+Multiframe+ or \verb+vvvSource+. Many of the tables 
are set up for individual programmes (e.g. an individual Public Survey), such as
\verb+vvvSource+ for the VVV, and \verb+videoSource+ for VIDEO. When we are 
discussing generic properties of ``Source'' tables, we will abbreviate them as 
\verb+Source+, rather than using programme-specific names. Individual columns 
within tables are referred to using a bold font: e.g. {\bf multiframeID} or 
{\bf aperMag3}.

\subsubsection{Metadata tables}

The following tables record metadata about images:

\begin{itemize}
  \item \verb+Multiframe+: This contains the main primary header keywords from
  the VISTA images and some additional derived quantities that are calculated
  for each multi-extension image. 
  \item \verb+MultiframeDetector+: This contains the main extension header
  keywords from the VISTA images except for astrometry related keywords and
  some additional derived quantities that are calculated for each detector.
  \item \verb+CurrentAstrometry+: This contains the astrometric related
  extension keywords and some additional derived astrometric quantities.
  \item \verb+MultiframeEsoKeys+: This contains subsidiary primary header
  keywords that are stored in the hierarchical ESO format (\verb+HIERARCH ESO+).
  \item \verb+MultiframeDetectorEsoKeys+: This contains subsidiary extension
  header keywords that are stored in the hierarchical ESO format.
  \item \verb+AstrometricInfo+: This includes additional derived
  astrometric properties, for OB frames (i.e. those created in a single
  observing block) used in multi-epoch surveys: the half-spaces (see
  \S~\ref{sec:halfspace}) for each edge of each frame and small offsets that can be applied to the frames to improve the astrometric precision.
\end{itemize}

\subsubsection{Catalogue data tables}
The following are the catalogue data tables used in the VSA. There is a
different table for each programme, so they will each start with the programme
acronym:

\begin{itemize}
  \item \verb+Detection+: This contains the extracted sources for each
  science stack detector frame in \verb+MultiframeDetector+ (individual frame in
  a multiframe): the raw extraction attributes from the original FITS table, the calibrated positions and magnitudes and a few other
  derived quantities.
  \item \verb+Source+: This is a merged filter catalogue from the deepest images
  in each pointing, and is made ``seamless'' \citep{WSA} to allow the user to
  find the most complete set of unique sources in the programme.
  \item \verb+SynopticSource+: This is a merged filter catalogue made from 
  detections in contemporaneous images. This is useful if colours of
  variable stars are needed. Only those programmes designed to have contemporaneous
  colours will have a \verb+SynopticSource+ table. 
  \item \verb+Variability+: This contains statistics for the light-curves of
  sources in multi-epoch programmes, allowing selection of variables based on
  different statistical quantities. \verb+VarFrameSetInfo+ is a useful
  supporting table for \verb+Variability+ and includes the fitted noise
  functions for each pointing.
\end{itemize}

\subsubsection{Linking tables}
The following are tables that link different types of catalogue data or metadata:

\begin{itemize}
  \item \verb+MergeLog+: For each pointing this lists the image frames in each
  filter from which the extracted detections were merged together to form the
  sources in the \verb+Source+ table. These are the deepest frames in each
  pointing.
  \item \verb+SynopticMergeLog+: For each pointing at each epoch, this lists the
  image frames in each filter from which the extracted detections were merged
  together to form the sources in the \verb+SynopticSource+ table.
  \item BestMatch tables: These link the sources in the \verb+Source+ table to
  each epoch detection in multi-epoch surveys, to match epochs for light-curves.
  There are \verb+SourceXSynopticSourceBestMatch+ tables for ``contemporaneous''
  filter data and \verb+SourceXDetectionBestMatch+ tables otherwise.
  \item Neighbour tables: These are simple tables containing all sources from the master table
  matching sources from the slave table within a specified radius. These can
  used for multiple purposes, such as to link with external surveys, 
  e.g.~\verb+vhsSourceXDR7PhotoObjAll+ links the VHS \verb+Source+ table to the 
  Sloan Digital Sky Survey Data Release 7 \verb+PhotoObjAll+ table.
  \item \verb+TilePawPrints+: This links tile image detections to the detections
  from pawprint images that make up the tile.
  \item \verb+Provenance+: This links image frames to their components, e.g.~a
  deep stack to each epoch stack frame that went into it, or an epoch stack
  frame to the raw images. 
  \item \verb+ProgrammeFrame+: This assigns image data to a programme and the
  programme requirements and is very important for programme curation. The same
  frame could be used in multiple programmes, for instance different PI 
  programmes with the same PI in different semesters or an all hemisphere
  release containing data from VHS, VVV, VIKING and VMC. 
\end{itemize}

\subsubsection{External catalogues}
The scientific goals of surveys tend to require external data (e.g.
from surveys on other telescopes / instruments at different parts of the
electromagnetic spectrum), in addition to data from VISTA itself. To support
those analyses, the VSA contains copies of catalogues from a number of external
surveys, which are listed in the online schema browser. The 
list of these is updated in response to requests from the survey consortia, and 
new cross-match neighbour tables are added for different programmes, as these 
external surveys become available. The online documentation explains how these 
cross-neighbour tables can be used to perform effective cross-catalogue queries.

\subsubsection{Curation tables}
\label{sec:curTab}
As mentioned above, the operations of the VSA are database-driven once the 
original MEFs have been ingested, with processing steps and data product 
provenance recorded automatically in the database. The VSA contains,
therefore, a large number of tables that drive, and are derived from, these 
curation tasks. Many of these are only of relevance to the VSA operations 
team, the following list do contain some pertinent information for users of the 
VSA:

\begin{itemize}
  \item \verb+Programme+: Basic programme information. This includes the
  programme dependent information used to create the SQL schema which
  drives most curation tasks.
  \item \verb+RequiredTile+: The current expected tile product pointings and
  matching tolerance. In the case of VIDEO, for which we ingest mosaics provided
  by the survey team, the relevant table is \verb+RequiredMosaic+.
  \item \verb+RequiredNeighbours+: lists which neighbour tables that join
  surveys have been created and what are the matching radii. 
  \item \verb+PreviousMFDZP+: The photometric calibration history of each image
  extension. MFDZP is a contraction of \verb+MultiframeDetector+ zeropoint. 
\end{itemize}

\section{Differences between WFCAM and VISTA Science Archives}
\label{sec:differences}

While the design of the WSA was developed with ultimate application to VISTA in mind,
there are some differences between the WSA and VSA structures. 

\subsection{Tile and pawprint information in the VSA}
In the VSA, we store catalogues from both the pawprints and tiles in the
detection tables for each survey (e.g.~\verb+vhsDetection+ for the VHS survey).
The tile catalogues are needed to produce uniform catalogues to the full depth 
of each survey. However, the astrometric solution in tile catalogues is not
quite as good as that in pawprint catalogues because the distortion is not as 
well represented by the tangent plane (TAN) projection which tiles are 
projected onto, as it is in the zenithal polynomial projection (ZPN) that can be 
used for the pawprints, \citep[see][]{project}. Saturated stars also have better
photometry in the pawprint catalogues. Producing pawprint catalogues does not 
add any additional overhead, since they must be produced as part of the production of tiles to allow the pawprints to be aligned
correctly before mosaicking.

Having both tiles and pawprints has created the need for multiple layers of
products and more complicated archive curation infrastructure (see 
\S~\ref{sec:infra}) to keep track of these and allow them to be used together. 
It also means that stack requirements need an additional constraint, the offset 
position. Each stack that goes into a tile has a different offset position, 0-5,
which is a function of the difference (in arcseconds) of the centre (optical
axis) of the pawprint from the centre of the tile. These offsets are stored as 
{\bf offsetX}, {\bf offsetY} in the \verb+Multiframe+ table. The offset position
is not the same as the {\bf offsetID} in \verb+Multiframe+, which is simply the order that the offset was observed in
and may differ in relative position on the tile from one epoch to the next (i.e.
the order in which the pawprints are executed can be chosen by the observer but
the relative positions of the 6 pawprints are fixed in the OBs currently
allowed). However, the {\bf offsetPos} always refers to the same part of the
tile. There can be considerable overlap between two pawprints, from different 
parts of two different tiles, but they will not be stacked together.

\subsubsection{The Tile-PawPrint matching tables}
\label{tilePaw}

To link the tile and pawprint catalogues together, we have
created two extra tables: \verb+TileSet+ and \verb+TilePawPrints+, which 
match each detection in a tile catalogue with detections at the same position
in the pawprint catalogues. These tables are survey specific, so VHS, which has
detections in \verb+vhsDetection+ will have tables \verb+vhsTileSet+ and
\verb+vhsTilePawPrints+ to link the tile and pawprint catalogues. 
\verb+TileSet+ and \verb+TilePawPrints+ are designed along the lines of 
\verb+SynopticMergeLog+ and \verb+SynopticSource+ \citep{Crs09}: \verb+TileSet+
links the frames together using the multiframe identifiers for the tile and
pawprints, and \verb+TilePawPrints+ links the detections using the extension
numbers (i.e. detector number) and sequence numbers (i.e. order that object was
extracted in the frame). The \verb+TilePawPrints+ table is deliberately as
narrow as possible, and simply includes the necessary linking information, with no
additional attributes such as magnitudes, since it is expected that it will
always be used to link, and could be used with a whole variety of attributes. By not including magnitudes, we also reduce the number of updates that are needed when
recalibrating the photometry. We have put some examples of linking tile
detections to pawprint detections in the VSA SQL
cookbook\footnote{http://surveys.roe.ac.uk/vsa/sqlcookbook.html\#TilePawprints}.

\begin{figure} 
\resizebox{\hsize}{!}{\includegraphics{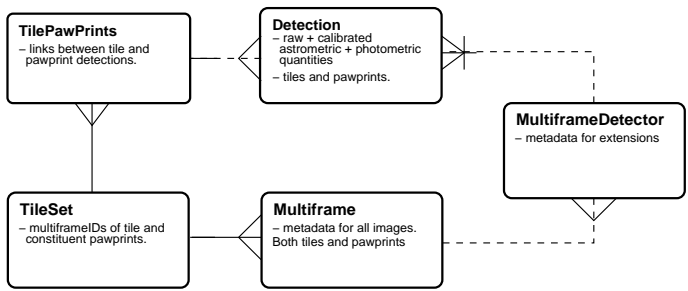}} 
\caption{The ERM for the tile-pawprint linking tables. {\texttt TileSet} links
the 7 multiframes (1 tile and its 6 constituent pawprints) in {\texttt
Multiframe}. Each row in {\texttt TileSet} links to many associated rows in the
{\texttt TilePawPrints} table, each of which refers to a different object. Each 
of these objects has between 1 and 7 measurements (detections) in the {\texttt
Detection} table. The complication comes from {\texttt TileSet} containing a mix
of tile and pawprint frames which cover different areas on the sky. The tile has
a single extension and the pawprints have 16 extensions in {\texttt
MultiframeDetector}.}
\label{fig:tilepawERM} 
\end{figure}

In Fig~\ref{fig:tilepawERM} we show an entity-relationship model (ERM) for these
new tables, showing how they are related to the current \verb+Multiframe+,
\verb+MultiframeDetector+ and \verb+Detection+ tables. The caption describes
the relationships.

Tile and pawprint detections are matched within a fixed radius of $0.8\arcsec$
in VISTA, which is approximately the average seeing and is many times the typical
astrometric error but less than the separation between easily-resolvable
neighbouring objects, so objects will only be matched to the same object on different images, not to 
neighbouring objects. The matching algorithm is the same as for sources in the 
\verb+Source+ table, see \cite{WSA}. Like \verb+MergeLog+, \verb+TileSet+ 
includes all the associated frames as a frame set consisting of the tile frame ({\bf
tlmfID}) and the 6 pawprint frames ({\bf o1mfID} -- {\bf o6mfID}) where {\bf o1} is the 
pawprint with {\bf offsetID}=1. \verb+TilePawPrints+ then contains the matched 
detections between the 7 frames, just like \verb+Source+ would have the matched 
detections between all the different filters. These tables should be used as 
linking tables to compare the tile detections to pawprint detections or even
pawprint to pawprint detections between offsets.

There is an important difference in the way that tile sets are produced that
causes some peculiarities to exist in the \verb+TilePawPrint+ tables that are
not present in the \verb+SynopticSource+ tables. Tile sets are merged from
frames of two different types -- a tile and six pawprints -- whereas synoptic
framesets are always merged from one frametype, either pawprints in WFCAM or
tiles in VISTA. In either case, each frame in the synoptic frame-set is similar
to each other and the matching condition is a combination of multiframe 
identity and extension number. Non-detections in a particular filter have a
default entry in the \verb+SynopticSource+ table: if there is no detection in
the J-band then  {\bf jSeqNum} is simply set to the standard integer default 
value (-99999999).

Since tile sets in \verb+TileSet+ are made up of a tile and six pawprints, with
the tile composed of and overlapping all 16 detectors in each pawprint, tile
sets cannot be matched by extension, but must be matched by multiframe alone. 
Therefore it is not possible to have entries in the tile set identified by {\bf
multiframeID} and {\bf extNum}, they must be identified by {\bf multiframeID} only, and the links in 
\verb+TilePawPrints+ must include both the extension number and sequence 
number. This makes the assignment of default rows more complicated. For
example, if there is no detection in a particular frame, such as pawprint
offset 1, we do not know off-hand -- without doing additional processing -- which extension the detection
should have been on (if any, since it may be in a gap between the detectors for
this offset). We should set the default row for this missing detection as 
${\bf o1mfID}={\bf multiframeID}$ of the pawprint, ${\bf
o1ExtNum}=-9999$, ${\bf o1SeqNum}=-99999999$. However, there is no equivalent 
row in the \verb+Detection+ table because the foreign key constraint between 
the \verb+Detection+ table and \verb+MultiframeDetector+ forbids this, since 
there are not rows in \verb+MultiframeDetector+ with a non-default {\bf
multiframeID}, and a default {\bf extNum}. Instead we set additional default
rows in \verb+TilePawPrints+, to have extension numbers equal to 2 and
default sequence numbers, so that they can match with default rows in the 
\verb+Detection+ table. 2 is used because it is the lowest number of a real
science extension, and can apply to both tiles and pawprints. These defaults are
extremely useful in queries that compare the photometry of tile-detected sources
in the merged-band catalogues to the pawprint detections. If a query compares
the photometry of the tile to each of the 6 pawprints, in most cases one or more
pawprints will not overlap with a particular tile detection; without these
defaults no row would be returned even if all 5 other pawprints had a match and
using these defaults, it is clear from the {\bf seqNum} value that it is a
non-detection. We must emphasize that an entry in \verb+TilePawPrints+, which
has a key ({\bf o1ExtNum}=2,  {\bf o1SeqNum}=-99999999) simply means that this
object was not detected in pawprint offset 1. It does {\bf not} denote that that
the object overlapped with extension 2 of pawprint offset 1: more than likely
it was from part of the tile which did not overlap with pawprint offset 1 at
all, although it could just be too faint to be detected in the pawprint.

Most \verb+TilePawPrints+ rows will be entries where the tile and two pawprints have
matched detections, some where the tile and just one (in the outer {\it
strips}) or three four, five or (infrequently) six pawprints have detections, 
some where the tile only has detections (usually at the faint end). Defaults are
added as above where no detection exists. Careful selection of what is default
and what is not will optimise the use of these tables. Any attributes in the 
detection tables can be compared in this way, although it is necessary to match 
a new instance of the \verb+Detection+ table for every frame in the table. Since
this is done via the primary key ({\bf multiframeID}, {\bf extNum}, {\bf
seqNum}), the joined SQL queries are very efficient.

\begin{figure} 
\resizebox{\hsize}{!}{\includegraphics{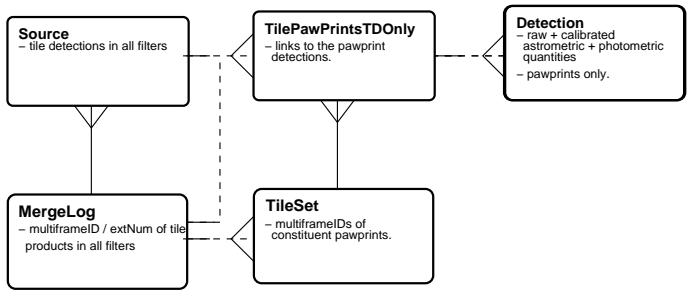}} 
\caption{The ERM for the tile-pawprint linking tables matched with the {\texttt
Source} table. Detections in the {\texttt Source} table come from tiles. In each 
{\texttt MergeLog} row there will be at least one non-default tile with a
maximum equal to the number of filters used in the programme. Each entry in
{\texttt MergeLog}, including defaults is matched to a {\texttt TileSet} and
each {\texttt Source} row to a similar number of rows in {\texttt 
TilePawPrintsTDOnly}, some of which may be default depending on whether there
was a detection in each observed filter. The relationship between {\texttt 
MergeLog} and {\texttt TilePawPrintsTDOnly} is necessary because the multiframe 
extension information in {\texttt MergeLog} is found in {\texttt
TilePawPrintTDonly}, not {\texttt TileSet}.}
\label{fig:sourPawERM} 
\end{figure}

To match tile detections in a \verb+Source+ table or \verb+SynopticSource+
table to the detections on the constituent pawprints, it is necessary to remove
the pawprint-only detections to leave a table with only the good tile detections
and necessary defaults. If a query retains the pawprint-only detections, they are
interpreted as defaults, so if a detection is missing in a particular filter it
will be matched to every set of pawprint detections that are not linked to a
tile-detection, a nonsensical result. To avoid this, we have created a view
\verb+TilePawTDOnly+ that can be directly matched in the same way as
\verb+TilePawPrints+. However, for very large datasets, such as the VVV, queries
work better if users use \verb+vvvTilePawPrints+ and add the necessary
constraints into the where clause, see example queries found in the SQL cookbook. 
The ERM for the matching of the \verb+Source+ table to the pawprint detections
is shown in Fig~\ref{fig:sourPawERM}. 

In the case of the VHS, which has a single epoch only, we have provided an
additional table \verb+vhsSourceXPawPrints+ which is a neighbour table between
\verb+vhsSource+ and the paw-print detections in \verb+vhsDetection+. This
simply matches all pawprint detections within a given radius to a source. A
typical query is shown at the bottom of the tile-pawprints section of the SQL
Cookbook. It is more difficult to do precision queries on particular offsets
or extensions as with \verb+TilePawPrints+, but it is possible to do faster
queries and may be preferable if only the pawprint data are required.

\section{Changes to image metadata}
\label{sec:im_meta_ch}

\subsection{ESO attributes}

VISTA data pass through an ESO quality control pipeline (whose modules
are provided by VDFS) in Garching before being ingested into the VDFS data
processing pipeline in Cambridge, while the VDFS-generated data products supplied to the ESO SAF must 
comply with ESO metadata standards. As a consequence, the VISTA data products 
present contain a quantity of standard ESO information not present in WFCAM data
products. For example, the headers of image files contain a number of ESO 
hierarchical FITS
keywords\footnote{http://heasarc.gsfc.nasa.gov/fitsio/c/f$\_$user/node28.html}. 
Those required for data processing, or judged to be scientifically useful, are 
propagated into the \verb+Multiframe+ or \verb+MultiframeDetector+ tables for 
keywords from primary or extension headers, respectively, while the remainder
are recorded in \verb+MultiframeEsoKeys+ and \verb+MultiframeDetectorEsoKeys+ tables.

Initial quality control occurs when the data is checked at the telescope and 
then at Garching to determine if the data was taken within the required
constraints. This results in additional quality control metadata created for
VISTA, which are included in the VSA, and which are not found in the WSA. These 
include OBSTATUS (``Completed'', ``Executed'', ``Aborted'', ``Pending'' and 
``Undefined''), and ESOGRADE (``A'', fully within constraints; ``B'', mostly -
90\% - within constraints; ``C''; ``D''; ``R'', rejected). If the OBSTATUS is 
not completed, the whole OB will be repeated later. Each OB is also quality
assessed more generally when processed in Cambridge.

There are also requirements for the files that are imported into the ESO Science
Archive Facility\footnote{http://archive.eso.org/cms/}. We generate
the following required keywords for the FITS files that are sent to the ESO
archives: ABMAGLIM, ABMAGSAT, MJDEND, TEXPSUM, respectively representing the
calculated $5\sigma$ magnitude limit of the point sources in the extension in AB mag, the
AB magnitude at which sources start to saturate, the end time (modified Julian
days) of the last exposure which went into the stack and the summed total
exposure of all the pawprints which went into the tile in seconds. These have
been added into \verb+MultiframeDetector+ as {\bf abMagLim} and {\bf abMagSat} and \verb+Multiframe+ as {\bf mjdEnd}, {\bf totalExpTimeSum}. Currently the AB saturated magnitude is only calculated when images are released to ESO, so all the values are default in the archive. The image pixel data have also been scaled and converted to 32-bit integer from
32-bit floating point, as a requirement for the ESO archive. The calculation of
the scaling parameter is shown in Appendix~\ref{app:scale}.

\subsection{Deprecations}
 
Deprecation codes in the \verb+Multiframe+, \verb+MultiframeDetector+,
\verb+Detection+ tables are used to control which frames are used where. The
different sub-surveys within UKIDSS followed the same deprecation policy, so it was possible to apply that uniformly across the whole WSA and not release any
deprecated data. For VISTA, however, the Public Survey teams have defined 
different deprecation criteria, so it is not possible to define an analogous 
uniform policy for the VSA. Instead, it has been decided to release all data 
(deprecated or not) but to define additional deprecation codes to indicate 
whether or not particular images have been omitted in the creation of higher 
order products: deep stacks, tiles or mosaics, \verb+Source+ tables, 
\verb+SynopticSource+ tables, neighbour tables or multi-epoch/variability
tables. The presence of most codes does imply exclusion from further use, but
the following codes, which are exclusive to the VSA, are more nuanced:

\begin{itemize}
  \item 50: The frame is good enough for single epoch measurements, but should
  not be used in a deep stack;
  \item 51: The frame has a problem with intermittency problem with channel 14
  (some early frames had this temporary issue in detector 6). During deep stack
  creation, channel 14 is set to zero weight in a temporary confidence image.
  \item 53: These are frames where the quality is marginal. Do not use the
  frames in deep stacks, or use the detections in the variability statistics,
  but do link them in the best match table, so the the survey team can do more tests to
  determine whether they are good enough to be used in future releases.
\end{itemize}

The following codes, are standard exclusion codes, but are not found in the WSA:

\begin{itemize}
  \item 55: Aborted OB, if the science team decided they want to deprecate based
  on OBSTATUS;
  \item 56: Deprecated because of poor ESOGRADE, if the science team decided
  they wanted to deprecate based on this;
  \item 58: Deprecated because the catalogue could not be ingested. This
  happened for many very dense fields in early processing versions, but these
  have since been replaced. We have included the code in this paper for the
  sake of completeness and for users who use older team releases that still
  contain data annotated with this code. If a similar situation arises in the future, more data may
  be deprecated with this value. 
\end{itemize}

\section{Changes to catalogue attributes}
\label{sec:cat_attr_ch}

Some of the catalogue attributes present in the VSA differ from those in the WSA, for several
reasons. Differences in the VISTA and WFCAM detector properties mean that
different effects need to be flagged, while the observing strategy changes described in 
\S~\ref{sec:tile_pawprint} led to requirements for different information. We have also 
introduced additional attributes in response to user demand and in the light of enhancements
we have made, especially in the treatment of multi-epoch data.

We describe all these changes in the remainder of this section.

\subsection{VIRCAM detector properties}
\label{detectors}

The Raytheon VIRGO detectors used in VIRCAM have no detectable crosstalk and
much lower persistence than the Rockwell Hawaii 2 detectors used in WFCAM. For this
reason, detections do not have their quality bit-flags set for crosstalk
\citep[bit 19 of the post-processing error bit flags -- {\bf ppErrBits},
see][]{WSA} in the VSA, which reduces processing time. However, the VIRCAM detectors have a narrower dynamic range, 
with non-linearity and saturation occurring at lower flux levels.
Non-linearity is calibrated out as part of the VDFS pipeline at CASU, and 
WFAU have applied a saturation correction \citep{satur}, where necessary, to the
photometry in the VSA --- note that users wishing to extract photometry for 
objects brighter than m~$\sim13$ (though this limit is survey dependent
according to the DIT and filter used in the OBs) should use pawprint detections
rather than tile detections since the corrections for saturation are more
accurate for pawprints.

On the top half of detector 16, the quantum efficiency (QE) varies on short
timescales making flat fields inaccurate. This is particularly
noticeable at short wavelengths ($\sim1\mu$m e.g.~Z and~Y). Since tiles are
produced from 6 pawprints, each with 16 detectors, we still create a tile even if
one or more of the constituent detectors has problems. Each tile pixel comes 
from up to 6 pawprints, which may have different PSFs (this is why
tiles are `grouted' as described earlier). Tile detections
that come partly from detector 16 in one or more pawprints are flagged using the post-processing error
flag ($\bf ppErrBits$) bit 12, so that users can select a data set without 
these detections or with them, whichever they prefer. Many of these detections 
have a low average confidence. We also added a flag for low confidence 
detections, bit 7.

Occasionally a particular pawprint detector is deprecated for one of several
reasons \citep{WSA}, e.g. poor sky subtraction, bad channels, detector was not
working correctly at the time of observation. The confidence of a deprecated
detector is set to zero when making the tile, and this produces poorly defined 
extractor values (infinities and not-a-number), which are ingested as defaults into the database. These tile detections are flagged with bit 24.

The two {\it strips} at the top and bottom of the tile have half the
exposure time of most other parts of the tile. We flag these with bit
23. This is partly for the users and partly so that these detections do not become primary
sources in the \verb+Source+ table of the survey if they overlap with a full
exposure region of another tile.

The list below is a summary of the new detection quality bit-flags developed for
VISTA tile detections.
\begin{itemize}
  \item Bit 7, Low average confidence ($<80$) in default aperture. 
  \item Bit 12, Source image comes partly from detector 16
  \item Bit 23, Source lies within a strip of the tile that has half the
  exposure of most of the tile.
  \item Bit 24, Source lies within an underexposed region due to a missing
  or deprecated detector 
\end{itemize} 

These tile flags have not been applied to catalogues from external mosaics
created for VIDEO, since their processing is quite different.

\subsection{Changes to attributes in {\texttt Detection} tables}

The VDFS extractor~\citep{CASU}, which generates the raw catalogue parameters
for all the VSA data \citep[apart from VIDEO mosaic catalogues, which are 
extracted using Source Extractor;][]{Btn96} has had a few modifications so that 
there are slightly different output parameters for VISTA than WFCAM. In the 
original FITS catalogues produced by the VDFS extractor the {\it 
Parent$\_$or$\_$child} column ({\bf deblend} in the WSA \verb+Detection+ tables)
has been replaced by {\it average$\_$conf} (stored as the {\bf avConf} in the 
VSA \verb+Detection+ tables), while the Hall radius, Hall flux and Hall flux
error have been replaced by a half-light radius ({\bf halfRad}) and flux and
flux error ({\bf halfFlux}, {\bf halfFluxErr}) within an aperture twice the
half-light radius.

\subsubsection{Modified Julian Day}

In WFCAM, we used the {\bf mjdObs} in \verb+Multiframe+ as the time for each
observation, which was used in light-curves. This attribute is inadequate in
VISTA though, since tiles are made from overlapping pawprints which each have 
different mean observation times and each tile detection may come from a
different combination of pawprints. This is particularly important for surveys 
such as the VMC \citep{VMC}, which require a significant fraction of an hour of
integration time to reach the required depth at each epoch, but are looking
for variables with periods of a few hours. In this case an accurate measurement
of the mean observation time is fundamental to the science. 

We have now added a new attribute {\bf mjd} into the detection tables. This is
the standard Modified Julian Date (MJD, in double precision days since midnight
on Nov 17th 1858) and is calculated detection by detection in the case of tiles
or extension by extension for pawprints. During `grouting', the average MJD of
each tile detection is calculated as the weighted average (weighted by the 
average confidence in a $1\arcsec$ radius aperture: aperture 3) MJD of the
different pawprints that contributed to the tile detection. In the FITS file, 
this is expressed as a 4-byte floating point value in minutes from the beginning
of the day of the observation, as column {\it MJDOff}, with the beginning of the
day given in the header as {\it MJD$\_$DAY}. We have also calculated the mean 
MJD for each pawprint detector and added this to \verb+MultiframeDetector+ as 
{\bf mjdMean}. This is the value that becomes the {\bf mjd} (8-byte double
precision) value in the detection tables for pawprints, not the {\bf mjdObs},
which is the start time of the observation.
 
\subsubsection{Half-light radii}

As well as ingesting detection attributes calculated by the VDFS extractor
produced in the FITS catalogue output, other attributes are calculated by the archive
curation software. These include several half-light radius measurements, based 
on the aperture fluxes and the Petrosian flux, using the same method discussed 
in \cite{SLC09}. These new attributes are: 

\begin{itemize} 
  \item {\bf hlCircRadAs}, the half-light circular radius in arcseconds.
  \item {\bf hlCircRadErrAs}, the error in the half-light circular radius in
  arcseconds.
  \item {\bf hlGeoRadAs}, the geometric mean between the half-light radius along
  the semi-major axis and the half-light radius along the semi-minor axis in
  arcseconds.
  \item {\bf hlSMnRadAs}, the half-light radius along the semi-minor axis in
  arcseconds.
  \item {\bf hlSMjRadAs}, the half-light radius along the semi-major axis in
  arcseconds.
  \item {\bf hlCorSMnRadAs}, the half-light radius along the semi-minor axis
  corrected for seeing, in arcseconds.
  \item {\bf hlCorSMjRadAs}, the half-light radius along the semi-major axis
  corrected for seeing, in arcseconds.
\end{itemize}

The algorithms used to calculate the above attributes are given in
Appendix~\ref{app:hlr}.

\subsubsection{Magnitude corrections}

Photometric calibrations, derived by the VDFS pipeline at CASU
(Hodgkin et al. 2012, in
preparation\footnote{http://http://casu.ast.cam.ac.uk/surveys-projects/vista/technical/photometric-properties}),
are applied in the archive curation software. As mentioned in \S~\ref{detectors}, we now include a saturation correction to the pipeline produced magnitudes of stars flagged as potentially saturated.  We
decided to include explicit columns that contain this and other source dependent 
corrections (those that are not simply field dependent), making it easy for 
users to understand and apply the corrections themselves. 

The current corrections that are applied to the magnitudes by WFAU in the VSA
are:

\begin{itemize}
  \item {\bf illumCorr}, the illumination or scattered light correction that is
  calculated and provided by CASU for fields on a month by month basis.
  \item {\bf distortCorr}, the radial distortion correction, which depends on
  the distance from the optical axis and the filter only.
  \item {\bf saturatCorr}, the saturation correction, that is added to
  the 1~arcsecond radius aperture magnitude ({\bf aperMag3}) of bright stars only
  (those that are flagged as potentially saturated).
  \item {\bf deltaMag}, the sum of the exposure time correction
  ($2.5\log_{10}{\bf expTime}$), the atmospheric extinction correction
  ($(0.5({\bf amStart}+{\bf amEnd})-1){\bf extinctionCat}$), the illumination
  correction and the radial distortion correction. The saturation correction is not included,
  because it only applies to {\bf aperMag3}. A user can calculate their own
  magnitudes on the VISTA photometric system for objects by measuring a flux
  in any way they like and applying the zero-point and adding {\bf deltaMag}.
\end{itemize}

The aperture corrections are not included since they are only applied to
specific magnitudes and are the same for all objects on one detector. The values
for these are included in the \verb+MultiframeDetector+ table. 

WFAU had several requests for aperture magnitudes without the point-source
aperture correction (i.e.~for extended sources). Therefore we have included
these values for the 7 apertures (1--7) for which aperture corrections have been
applied as standard. These are named {\bf aperMagNoAperCorr1}, {\bf
aperMagNoAperCorr2} to {\bf aperMagNoAperCorr7}. These magnitudes can be used
for extended sources if required.

\subsection{Changes to attributes in the {\texttt Source} tables}

Several of the attributes in the \verb+Detection+ table have been 
propagated through to the \verb+Source+ table or \verb+SynopticSource+ table.  
Of the new \verb+Detection+ attributes, we have
propagated {\bf hlCorSMjRadAs} and the non-aperture corrected aperture
magnitudes into the \verb+Source+ table. We do not propagate {\bf mjd}
though, since the sources in the \verb+Source+ table come from the deepest data, 
stacked across multiple epochs where the time of observation is not particularly
useful. Since the \verb+SynopticSource+ matches data with a specific epoch, and is
particularly useful for variability work on point sources,
{\bf mjd} is propagated but {\bf hlCorSMjRadAs} and {\bf aperMagNoAperCorr[1-7]}
are not.

We still only produce one \verb+Source+ table, from the highest (most processed
image, i.e. tile if the survey contains both tiles and stacks, mosaics if the
survey contains these) product layer (see \S~\ref{sec:infra}). Producing one for
tiles and one for pawprints breaks the idea of a single master source list \citep[see][]{Crs09}. Instead, we have a master source list produced from tiles which is linked to pawprints using the
\verb+TilePawPrints+ table, see \S~\ref{tilePaw}. 

\section{Infrastructure}
\label{sec:infra}

There have been several changes to the science archive infrastructure that
improve curation of the surveys, but can also be useful for scientists who 
want to make the best use the VSA. Some of these changes have been incremental 
and have been documented in \cite{Cll09,Crs09,Crs10}. The main changes to the 
VISTA Public Surveys from the UKIDSS Public Surveys are listed below:

\begin{itemize}
  \item Automatically set up all the requirements, the database schema and
  curation tables contents using available data and basic programme properties 
  from the \verb+Programme+ table. This was also done for the UKIDSS-DXS and 
  WFCAM PI programmes.
  \item Manage multiple layers of products: pawprints, tiles and mosaics,
  including external products automatically.
  \item Have a more sophisticated setup for multi-epoch products, specified by
  the {\bf synopticSetup} string in \verb+Programme+.
\end{itemize}

We set up all the requirements, the database schema and curation tables contents
for a survey, when we start preparing a static release (e.g.~VHS-DR1), using a 
combination of the programme requirements in the \verb+Programme+ table and the 
available data. We have made the infrastructure and processing the same for all
surveys, unlike in UKIDSS where the wide shallow surveys (GPS, GCS and LAS) 
were processed differently from the deep surveys (DXS and UDS). This makes it 
easier for the operators who run tasks, and makes it much simpler if programmes 
evolve in the future. For instance, if the VHS decided to add in a second epoch 
in any filter, this would be automatically accommodated.

\subsection{Stack, tiles and mosaics} 
 
Requirements for stack, tile and mosaic products are set up by grouping the
data into different pointings by position, position angle and, in the case of
pawprints, offset. The requirements for a particular release
are stored in \verb+RequiredStack+, \verb+RequiredTile+,
and \verb+RequiredMosaic+. The stacking software uses the definitions to
create the deepest stack possible for each product in \verb+RequiredStack+ from
the pawprints. The tiling software creates tiles at each location in
\verb+RequiredTile+ using these stacks, so the tile requirements must be linked
to the pawprint requirements.

The different layers of products (pawprints, tiles and mosaics) can be
linked to each other using the \verb+ProductLinks+ table: e.g.~tile productID 1
in \verb+RequiredTile+ in the VHS may be composed of pawprints with productIDs
1, 3, 5, 7, 9 \& 11 in \verb+RequiredStack+. \verb+ProductLinks+ links the
requirements whereas \verb+Provenance+ links the image metadata from the actual
files. From one release to another the values in \verb+ProductLinks+,
\verb+RequiredTile+ and \verb+RequiredStack+ may stay the same (although this is not guaranteed),
but a product which initially contained 2 epochs worth of image data may be
replaced by one containing 5 epochs worth of image data and will therefore 
link to different multiframes in \verb+Multiframe+ and \verb+Provenance+.
 
External (made outside VDFS) products, e.g.~VIDEO (or UKIDSS-UDS in the WSA)
mosaics, which are created by the survey team and imported into the VSA, are 
set up via the \verb+ExternalProducts+ table, which contains the programme, 
product type, release number and information about who created the mosaic.

The required products and the actual image frames are now linked to each other
via the \verb+ProgrammeFrame+ table which includes {\bf programmeID}, {\bf
productID} and {\bf releaseNum} and links to the image metadata tables via {\bf
multiframeID}. The release number for products is a running number from when WFAU
first started producing releases for the science teams, so the products in the
first public releases of the VISTA Public Surveys have a variety of release
numbers depending on the programme. In VISTA, the programme translation for
each incoming FITS image is more complicated than for WFCAM, and the programme
matching algorithm uses a combination of the {\bf HIERARCH ESO OBS PROG ID},
{\bf HIERARCH ESO OBS NAME} and {\bf HIERARCH ESO DPR CATG} header keywords.

\verb+ProgrammeFrame+ is essential for
keeping track of what images are related to each requirement. This makes it 
much easier for scientists and VSA support staff to keep track of what has been
created and whether anything is missing. This infrastructure is crucial for the
automated curation \citep{Cll09} of VSA products, where decisions are made 
about what tasks need to be run based on the requirements and what has already
been completed.

When OB frames are recalibrated in multi-epoch programmes, the OB tiles are
compared to the deep tiles and the zero-points adjusted accordingly. A change to
the zero-point of the tile is propagated to the constituent pawprints. The code to
propagate the zero-point differences was not developed until very recently, so
most datasets in the first release will not include this propagation; at the time of writing only the
VVV dataset will include this. The pawprint zero-points for the other multi-epoch
public surveys (VIKING, VMC and VIDEO) will be correctly updated for all
recalibrations in the data releases that contain data from ESO semester P87 and
beyond.

\subsection{Multi-epoch tables}

We have introduced a new string attribute into the \verb+Programme+ table,
called {\bf synopticSetup}, to control the production of more than one Best
Match table, i.e.~both a \verb+SourceXDetectionBestMatch+ and a \verb+SourceXSynopticSourceBestMatch+. In surveys such as the VVV, many scientists would like colour information for
variable stars, so the colours must come from near-contemporaneous observations. This
information is in the \verb+SynopticSource+ table whereas the colours in the
\verb+Source+ table come from the deepest images which are stacked from
several epochs of data and are certainly not contemporaneous. However, this
survey will take many tens of epochs, mostly in one filter, $K_s$, so a 
\verb+SynopticSource+ table that covers the full time range would be 
inefficient - the $Z$, $Y$, $J$ and $H$ band attribute columns would contain 
mainly default values. A more efficient way is to specify the
\verb+SynopticSource+ over a short time range and specify that the statistics
in the \verb+Variability+ should come from data in
\verb+SourceXDetectionBestMatch+, which covers the whole time range. It is still
necessary to link the \verb+SynopticSource+ table with all the other tables, so
a \verb+SourceXSynopticSourceBestMatch+ table is required too.The {\bf
synopticSetup} attribute is a string, with the following value in the VVV: 
\verb+BOTH:VAR-UNC:COR,SV,P87+, which can be parsed to give the following
information: create both Best Match tables; use the uncorrelated 
(\verb+SourceXDetectionBestMatch+) when calculating the variability statistics, 
and only use data between the beginning of the Science Verification period (SV)
and the end of ESO Period 87 (P87) in the correlated table 
(\verb+SourceXSynopticSourceBestMatch+).

\section{Other recent improvements to the VSA and WSA}
\label{sec:newFeatures}

In addition to the above changes necessary for processing VISTA data,
we have made various changes to improve overall curation of WFAU products.
These extend the database design described in \cite{WSA} and \cite{Crs09}.

\subsection{Improvements to multi-epoch data model and calculations}

\subsubsection{Creating the {\texttt BestMatch} tables}
\label{sec:halfspace}

In \cite{Crs09} \S~9.2, we discussed possible improvements to checking missing
observations to correctly create the \verb+BestMatch+ tables. One method that we
discussed possibly implementing was the half-space method \citep{Bud10}, which
we have now implemented. We define 16 half-spaces for each single epoch image.
These 16 half-spaces come in 4 sets, one 2 pixels outside each image edge, one 2 pixels inside each image
edge, one 2 pixels outside the edge of the jittered region where the exposure
time per pixel goes from the total exposure time to some fraction of it and one
2 pixels inside this edge. With 4 edges, this gives 16 half-spaces. We found
that defining a half-space using 3 points: the two ends of an edge and the
midpoint, the edge could usually be described to an accuracy of around one
pixel, so a 2 pixel margin each side would encompass all points which we were
unsure about. Each half-space is described using 4 numbers, a 3-D
Cartesian vector, normal to the plane of the half-space and a constant that
gives the offset of the plane from the centre of the sphere. All the half-space
information is stored in a new table, \verb+AstrometricInfo+ (one for each
multi-epoch programme, e.g. \verb+vmcAstrometricInfo+), to help with the
curation of multi-epoch data. As well as the half-space information, we have
included place-holder columns for attributes to describe small adjustments to the astrometric solution
of each image that will improve fitting to the proper motion, when we start
calculating proper motions in VISTA data \citep[see ][ for the description of
the method used for wide area UKIDSS surveys]{Coll_PM}. 

The half-spaces are used to check frames which do not have an expected match to
a primary source in the \verb+Source+ table, \citep[see][\S~9.2]{Crs09}. If
there is no detection, this may be for one of several reasons: the frame does
not overlap with the part of the sky containing the source; the source is within
the jitter regions, where the integration time is less and there may be a
gradient in the integration time across the object; the source is too faint to
be detected on a single exposure; the source is usually bright enough, but has
faded below the detection threshold; the source is blended with another; the
object has moved sufficiently far from the expected position.

Using the half-spaces allows us to flag the first two possibilities. Checking
whether a detection should be within the image or within the jitter section is 
trivial and each calculation is extremely quick. Most importantly, since the 
half-space describes an edge accurately within a pixel or two, very few objects 
need the more careful test that use the WCS to calculate the exact position of
the object on the frame. Using the half-spaces we are able to reduce the
number of slow tests to only those objects within two thin strips, each 4 pixels
wide, at the image edge and the edge of the jitter region. The half-space
information is stored so that archive users can use it too, to rapidly determine
whether an object is within the frame.

\subsubsection{Expected noise model}
We have made changes to the calculation of the expected noise.
The expected noise is still based on fitting a function, in this case, a
Strateva function to the RMS versus mean magnitude data \citep[see ][]{Crs09}.
In early team data releases (before version 1.1 data was released), the expected
noise was simply the value of this function at the mean magnitude of the source.
However, we found that the actual magnitude limit was often quite a bit brighter than the expected magnitude limit, especially when the field is very dense, see
Fig~\ref{fig:expRms}. When this happens the RMS versus magnitude plot turns over
and simply using the fit will underestimate the RMS; indeed sometimes the 
expected RMS will be negative. To mitigate against this, we have made the
following changes: 

\begin{figure}
\resizebox{\hsize}{!}{\includegraphics{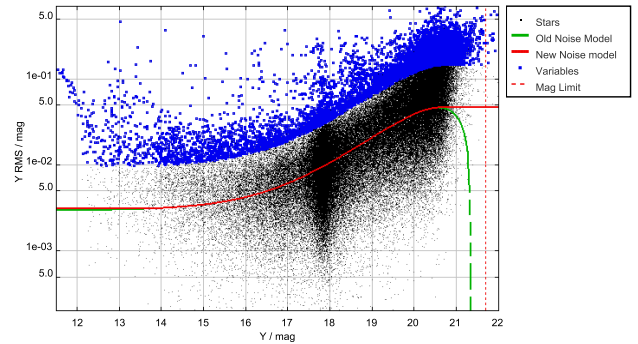}}
  \caption{Magnitude-RMS plot of a dense pointing in the
  VMC, where the fitting function turns over before the expected magnitude limit (dotted
  vertical line). The old expected RMS  as a function of magnitude is plotted as
  the dashed line, and the new expected RMS as a function of magnitude is
  plotted as the solid line. Outside of the fitting range, or beyond the
  turnover point, the RMS is a constant. Non-variable stars are shown as dots,
  and variables are shown as open squares. At the bright end, $Y<12.5$ mag, the
  RMS increases due to saturation effects, but this has not been included in the model yet.}  	
\label{fig:expRms}
\end{figure}

\begin{itemize}
  \item Calculate the turnover point: the maximum RMS as a function of
  magnitude if the function does turn over.
  \item For all magnitudes fainter than the turnover point or the maximum range
  of magnitudes in the fit (whichever is brightest), set the expected RMS to the
  value at this point, see Fig~\ref{fig:expRms}.
  \item For all magnitudes brighter than the minimum range of magnitudes in the
  fit, set the expected RMS to the value at this point. 
\end{itemize}

The astrometric fit has also been updated. Instead of calculating a
simple clipped mean, we calculate a weighted mean position. We
calculate the expected astrometric noise in each filter, in the same way as we 
calculate the expected photometric noise, using the angular separation between 
each epoch position and the median absolute deviation clipped median position
for stationary stars and fitting a function to these median values as a function
of magnitude. This function describes the locus of the astrometric uncertainty
for non-moving point sources, much as the equivalent fit for the magnitude RMS
described the photometric uncertainty as a function of magnitude for 
non-variable point sources. This calculated uncertainty as a function of
magnitude will be used to weight the position.

A particular pointing may only be observed once in one filter, and in some
programmes a filter is only observed once in each pointing, so it is often not
possible to fit a noise model for each pointing and filter. For photometric
variability statistics, this is not a problem, since all the values are default
if there is only one epoch, but when it comes to the astrometric fit, we would
like to use all the data in all filters together, to improve the fit. 

To estimate the errors on these frames, we calculate a default noise model in each filter which
has a calculable noise in at least one pointing. We take all the calculated
noise models, and calculate the mean RMS of these models in each of a set of
nine bins across the magnitude range, and fit the noise model to the mean RMSs. This
model is used in any pointing in this filter where there is only one epoch. 

For filters where there is only ever one epoch, we cannot directly measure the
noise as a function of magnitude, so we make the assumption that the behaviour
in any filter is similar to the others (which is born out by experience of
surveys where multiple observations are taken in all filters). We
expect that the limit reached at the bright end for the same DIT in the OBs will
be the same and that the increase in noise toward the faint end depends on the 
depths of the exposure, which depends on the total exposure time or expected 
magnitude limit. Moreover, any other differences, such as the effects of sky brightness or
residual non-linearity are likely to be a function of wavelength, so we choose
the nearest filter in wavelength that has enough epochs for a fit to be made to the RMS as a function
of magnitude. We take the default model in this filter and adjust for the
difference in expected magnitude limit, e.g. using the Strateva model, we will
calculate new values for $b$ and $c$ as follows

\begin{eqnarray*}
<\zeta(m)> & = & a+b\,10^{0.4m}+c\,10^{0.8m} \\
\Delta\,m & = & m_{\rm l}-m'_{\rm l} \\
a & = & a' \\
b & = & b'\,10^{-0.4\Delta\,m} \\
c & = & c'\,10^{-0.8\Delta\,m}, 
\end{eqnarray*} 

\noindent where $m_{\rm l}$ is the magnitude limit of frames in this filter and
$m'_{\rm l}$ is the magnitude limit of frames in the comparison filter.

The weighted mean position uses a 3$\sigma$ clipped weighted mean 
in each of the three Cartesian coordinates, and then converts back to equatorial
coordinates. The type of fit used is recorded in the \verb+VarFrameSetInfo+
table as {\bf motionModel}. The model described above is a static
weighted model: `wgtstatic'. When we have VISTA data over several years we
expect to fit for proper motion too.

\subsection{Improvements to the interface} 
 		 
The VSA proprietary and public release databases can be accessed and 
queried via the web-browser based interface. The various access methods 
allow users to perform SQL queries on the science ready tables; 
extract image cut-outs and download entire image and catalogue files. 
In addition public releases will be accessible under the Virtual 
Observatory (VO). Releases will be discoverable in the VO registries. 
A Table Access Protocol\footnote{http://www.ivoa.net/Documents/TAP/} (TAP) 
interface to each data release will allow users to perform SQL queries
using the Astronomical Data Query Language (ADQL). We already have 
partially compliant TAP services available for all our main data
holdings. They can be accessed through software like TOPCAT \citep{TOPCAT} and
VOExplorer, as well as through standard HTTP GET and POST commands. The 
services and their endpoints are registered on the major VO registries. We also
have conesearch and Simple Image Access Protocol 
(SIAP\footnote{http://www.ivoa.net/Documents/SIA/}) services available on  the
VO for the same list of datasets.

Fully compliant TAP services will likely be available by the end of 2012.

\section{Illustrative science examples}
\label{sec:queries}

\subsection{Colours of VHS point sources} 

The optical-infrared colour-colour plot is a powerful
classifier of different types of stars, with most stars 
lying along a narrow locus. However, extinction and poor photometry 
can widen this locus and prevent the separation of brown dwarfs, QSOs and
compact galaxies. The following selection will select point sources in the VHS,
which are matched to point sources in the SDSS and are not flagged for poor 
quality. We select the colours and positions, but only for stars in areas of 
low Galactic extinction.

\begin{verbatim} 
SELECT s.sourceID, s.ra, s.dec,
/* select position colour and magnitude 
information */ 
(sdPho.psfMag_g-s.jAperMag3) AS gmjPnt, 
jmksPnt, ksAperMag3
/* from vhsSource, SDSS DR7 PhotoObjAll, 
neighbour table */ 
FROM vhsSource as s, 
vhsSourceXDR7PhotoObjAll as x,
BESTDR7..PhotoObjAll as sdPho
/* join the tables */ 
WHERE s.sourceID = x.masterObjID AND 
x.slaveObjID=sdPho.objID
/* find matches within 2 arcsec */ 
AND x.distanceMins<=0.0333
/* that are nearest matches */ 
AND x.distanceMins IN (
/* sub query to find minimum distance 
for a match to this sourceID */ 
SELECT MIN(distanceMins)
FROM vhsSourceXDR7PhotoObjAll
WHERE masterObjID=x.masterObjID) 
/* select SDSS primary objects and stars */ 
AND x.sdssPrimary=1 and x.sdssType=6 
/* objects with no flags in VHS */ 
AND jppErrBits=0 AND ksppErrBits=0 
/* stars or probable stars in VHS */ 
AND mergedClass IN (-1,-2) 
/* Not default magnitudes in SDSS or VHS */ 
 AND sdPho.psfMag_g>0. AND s.jAperMag3>0.
AND s.ksAperMag3>0. 
\end{verbatim} 

\begin{figure} 
\resizebox{\hsize}{!}{\includegraphics{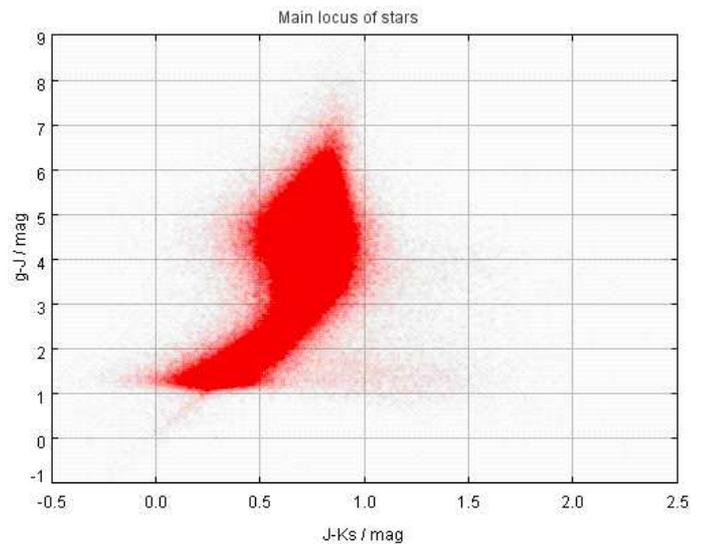}} 
\caption{The main locus of stars matched between the VHS and SDSS selected in 
 (g-J) versus (J-Ks).} 
 \label{fig:mainStellLocus} 
\end{figure}

These data can be plotted using TOPCAT \citep{TOPCAT}, and the main locus
can be found, as seen in Fig.~\ref{fig:mainStellLocus}. The gradient of the
locus in the colour-colour plot can be measured and then rare objects can be
further selected.
	 
\subsection{Flare stars} 
\label{sec:selFlare} 		 
 The following query selects objects that could be flaring stars or some type of
 cataclysmic variable. To do this, we select sources that have a minimum K$_s$
 magnitude that is at least 2 magnitudes brighter than the median magnitude and
 further, it has at least 2 measurements that are brighter than the median by
 0.5 magnitudes. This second constraint should remove sources with one point
 that has escaped flagging. We also want at least 5 good K$_s$ detections for
 a reasonable light curve. The following query was performed on the VVV survey.
 		 
 \begin{verbatim} 
 SELECT v.sourceID, s.ra, s.dec,
 /* select some useful attributes, pointing info,
number of observations, min, medium, maximum,
variable class, and star/galaxy class */ 
v.framesetID, ksnGoodObs, ksMinMag, ksMedianMag,
ksMaxMag, variableClass, mergedClass,
(ksMedianMag-ksMinMag) as ksFlareMag, 
COUNT(*) AS nBrightDetections
/* from vvvVariability and vvvSource */ 
FROM vvvVariability as v,vvvSource as s, 
vvvSourceXDetectionBestMatch as b, 
vvvDetection as d
/* first join the tables */ 
WHERE s.sourceID=v.sourceID AND b.sourceID=
v.sourceID AND b.multiframeID=d.multiframeID 
AND b.extNum=d.extNum AND b.seqNum=
d.seqNum AND
/* select the magnitude range, brighter than 
Ks=17 and not default. */ 
ksmedianMag<18. and ksmedianMag>0. AND 
/* at least 5 observations */
ksnGoodObs>=5 AND ksbestAper=5 AND
/* Min mag is at least 2 magnitudes brighter 
than median mag(but minMag is not default) */ 
(ksmedianMag-ksminMag)>2. and ksMinMag>0. AND 
/* Only good Ks band detections in same 
aperture as statistics are calculated in*/ 
d.seqNum>0 AND d.ppErrBits IN (0,16) AND 
d.filterID=5 AND d.aperMag5>0 AND 
d.aperMag5<(ksMedianMag-0.5) 
/* Group detections */ 
GROUP BY v.sourceID, s.ra, s.dec,
v.framesetID, ksnGoodObs, ksMinMag, ksMedianMag,
ksMaxMag, variableClass, mergedClass
HAVING COUNT(*)>2 
/* Order by largest change in magnitude first.*/
ORDER BY ksMedianMag-ksMinMag DESC 
\end{verbatim} 
 
\begin{figure} 
\resizebox{\hsize}{!}{\includegraphics{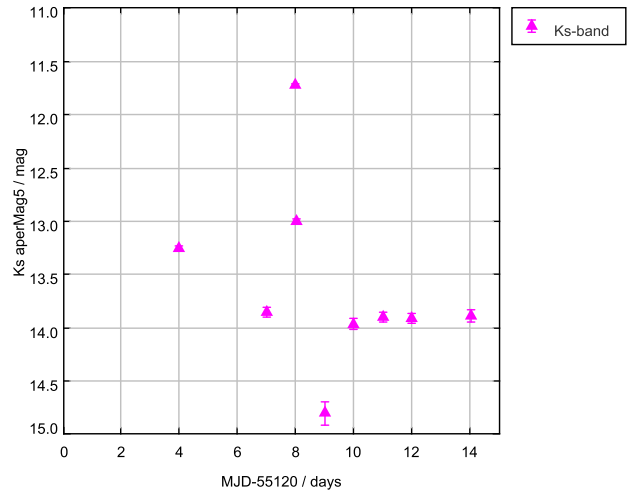}} 
\caption{The light curve of a flare star in the VVV selected using the query 
in \S~\ref{sec:selFlare}} 
\label{fig:flareStar} 
\end{figure}

We plot the Ks-band light curve of one of these objects in 
Fig~\ref{fig:flareStar}. The majority of the detections are $14^{th}$ magnitude,
but there is a flare of almost 2.5 mag followed by fading of 1 mag before the
star returns to $K_{s}=13.9$ mag. 
	 
\subsection{Global properties of VIKING-SDSS galaxies} 

The following selection uses IR photometry and sizes from VIKING combined 
with optical colours and 
redshifts -- both spectroscopic and photometric -- from SDSS. In this query we 
use neighbour tables to join VIKING and SDSS. We use the SQL command ``UNION'' 
to combine the query which matches to galaxies with spectra to the query for 
those with only photometric redshifts. Users who are worried about 
completeness can use the ``UNION'' command to combine further queries, such
as those that select for VIKING galaxies without SDSS matches or ones for SDSS 
matches but neither spectroscopic or photometric redshifts. The two combined queries 
must have the same number of columns, with the same names. In the cases where 
one query has columns which the other has no entry for (e.g. redshift, 
z) we can fill this column with a default number, just as we demonstrate below 
with the redshift status and spectroscopic type. 

\newpage 
\begin{verbatim} 
SELECT 
/* select information necessary to create
bi-variate brightness distribution,
extinction corrected Petrosian magnitudes
put into AB system and seeing corrected,
semi-major axis size, PLUS SDSS colours
and redshifts (spectroscopic and
 photometric) 
*/
s.sourceID,s.ra,s.dec,s.frameSetID, 
(s.hPetroMag-s.aH+fh.VegaToAB) AS hPetroAB, 
s.hHlCorSMjRadAs,(s.hPetroMag+ 
2.5*log10(2.*3.14159*s.hHlCorSMjRadAs* 
s.hHlCorSMjRadAs)-s.aH+fh.VegaToAB) AS
hSurfBright, (s.ksPetroMag-s.aKs+ 
fks.VegaToAB) as ksPetroAB, 
s.ksHlCorSMjRadAs,(s.ksPetroMag+ 
2.5*log10(2.*3.14159*s.ksHlCorSMjRadAs* 
s.ksHlCorSMjRadAs)-s.aKs+fks.VegaToAB)
AS ksSurfBright, dr7spec.objID as sdssID, 
((dr7spec.modelMag_u-dr7spec.extinction_u)- 
 (dr7spec.modelMag_g-dr7spec.extinction_g))
as umgModel,z,zErr,zConf,zStatus,specClass
/* from vikingSource, Filter (one for each
filter for VegaToAB), SDSS-DR7 neighbour
table, SDSS SpecPhoto table */ 
FROM vikingSource AS s,Filter AS fh,Filter
AS fks, vikingSourceXDR7PhotoObjAll AS xdr7, 
BESTDR7..SpecPhotoAll as dr7spec
/* First join tables, */
WHERE xdr7.masterObjID=s.sourceID AND
fh.filterID=4 AND fks.filterID=5 AND
dr7spec.objID=xdr7.slaveObjID AND 
 /* select VIKING primary sources matched to
 SDSS primary sources */
 (priOrSec=0 OR priOrSec=frameSetID) AND
sdssPrimary=1 AND
dr7spec.sciencePrimary=1 AND 
/* within 2" of nearest match */ 
 xdr7.distanceMins<0.03333 AND
 xdr7.distanceMins IN ( 
  SELECT MIN(distanceMins)
  FROM vikingSourceXDR7PhotoObjAll 
  WHERE masterObjID=xdr7.masterObjID AND
   sdssPrimary=1) AND 
/* for objects classified as galaxies or
probable galaxies in VIKING */ 
mergedClass IN (1,-3) AND 
/* h and ks size is 0.7<sma<=10. arcsec */ 
ksHlCorSMjRadAs>0.7 AND hHlCorSMjRadAs>0.7
AND ksHlCorSMjRadAs<=10.0 AND
hHlCorSMjRadAs<=10.0 AND
/* good quality data in VIKING h and ks */ 
hppErrBits=0 AND ksppErrBits=0 AND 
 /* ks extinction corrected AB mag < 20.5 */ 
 (ksPetroMag-aKs+fks.VegaToAB)<20.5 AND 
/* ra and dec range to restrict to where
SDSS is */ 
s.ra>100. AND s.ra<250. AND s.dec>-5. AND 
/* z>=0.002 */ 
 dr7spec.z>=0.002 
 /* Add in ones which do not have SDSS spectra
using UNION */ 
UNION 
SELECT 
/* select information necessary to create
bi-variate brightness distribution, 
extinction corrected 
Petrosian magnitudes put into AB system and 
seeing corrected, semi-major axis size AND SDSS 
matches to PhotoObj table and photoz table */ 
s.sourceID,s.ra,s.dec,s.frameSetID, 
(s.hPetroMag-s.aH+fh.VegaToAB) AS hPetroAB, 
s.hHlCorSMjRadAs,(s.hPetroMag+ 
2.5*log10(2.*3.14159*s.hHlCorSMjRadAs* 
s.hHlCorSMjRadAs)-s.aH+fh.VegaToAB) AS 
hSurfBright, (s.ksPetroMag-s.aKs+fks.VegaToAB) 
as ksPetroAB,s.ksHlCorSMjRadAs,(s.ksPetroMag+ 
2.5*log10(2.*3.14159*s.ksHlCorSMjRadAs* 
s.ksHlCorSMjRadAs)-s.aKs+fks.VegaToAB) AS 
ksSurfBright, dr7phot.objID as sdssID, 
((dr7phot.modelMag_u-dr7phot.extinction_u)- 
(dr7phot.modelMag_g-dr7phot.extinction_g)) 
as umgModel, photz.z as z,photz.zErr as zErr, 
-9.9999 as zConf,-9 as zStatus,-9 as specClass 
/* from vikingSource, Filter (one for each 
filter for VegaToAB), SDSS-DR7 neighbour 
table, */ 
FROM vikingSource AS s,Filter AS fh,Filter AS 
fks,vikingSourceXDR7PhotoObjAll AS xdr7, 
BESTDR7..PhotoObjAll as dr7phot, 
BESTDR7..photoz as photz 
/* First join tables, */ 
WHERE xdr7.masterObjID=s.sourceID AND 
fh.filterID=4 AND fks.filterID=5 AND 
dr7phot.objID=xdr7.slaveObjID AND photz.objID= 
dr7phot.objID AND dr7phot.objID NOT IN ( 
  SELECT dr7spec.objID
  FROM BESTDR7..SpecPhotoAll as dr7spec  
  WHERE dr7spec.objID=xdr7.slaveObjID AND  
  dr7spec.sciencePrimary=1) AND 
  /* select VIKING primary sources matched to  
  SDSS primary sources */  
  (priOrSec=0 OR priOrSec=frameSetID) AND  
  sdssPrimary=1 AND  
   /* within 2" of nearest match */ 
   xdr7.distanceMins<0.03333 AND  
   xdr7.distanceMins IN ( 
   SELECT MIN(distanceMins)  
   FROM vikingSourceXDR7PhotoObjAll 
   WHERE masterObjID=xdr7.masterObjID AND  
   sdssPrimary=1) AND 
   /* for objects classified as galaxies or  
probable galaxies in VIKING */ 
mergedClass IN (1,-3) AND 
/* h and ks size is 0.7<sma<=10. arcsec */ 
ksHlCorSMjRadAs>0.7 AND hHlCorSMjRadAs>0.7 AND  
ksHlCorSMjRadAs<=10.0 AND hHlCorSMjRadAs<=10.0  
/* good quality data in VIKING h and ks */ 
AND hppErrBits=0 AND ksppErrBits=0 AND 
/* ks extinction corrected AB mag < 20.5 */ 
(ksPetroMag-aKs+fks.VegaToAB)<20.5 AND 
/* ra and dec range to restrict to SDSS */ 
s.ra>100. AND s.ra<250. AND s.dec>-5. AND 
/* z>=0.002 */ 
photz.z>=0.002 
\end{verbatim} 

\begin{figure} 
\resizebox{\hsize}{!}{\includegraphics{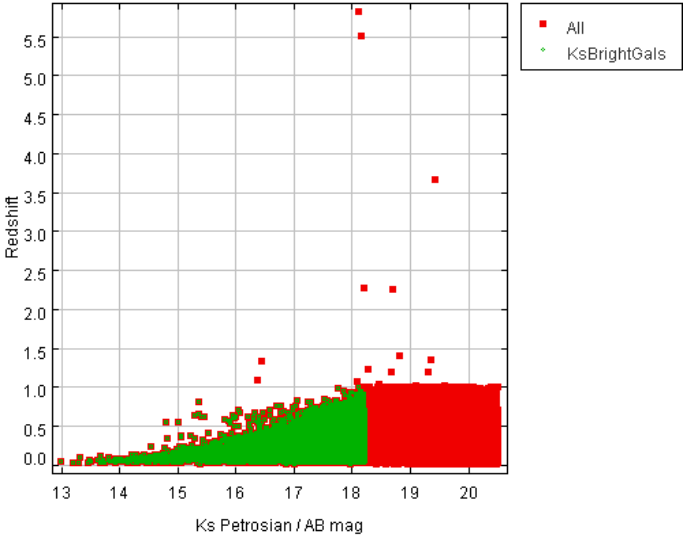}} 
\caption{Ks-band AB Petrosian magnitude versus redshift. This plot shows very 
few redshifts greater than $z=1$, shown in red. The galaxies with $z>1$ are 
likely to have spectroscopic redshifts, whereas the photometric redshifts are limited to 
$0<z<1$. Galaxies with $Ks_{AB}<18.2$ tend to have $z<1$, so we have selected
a sample in TOPCAT which have $Ks<18.2$ and $z<1$, which are shown in green.} 
\label{fig:KABvsRed} 
\end{figure}

\begin{figure} 
\resizebox{\hsize}{!}{\includegraphics{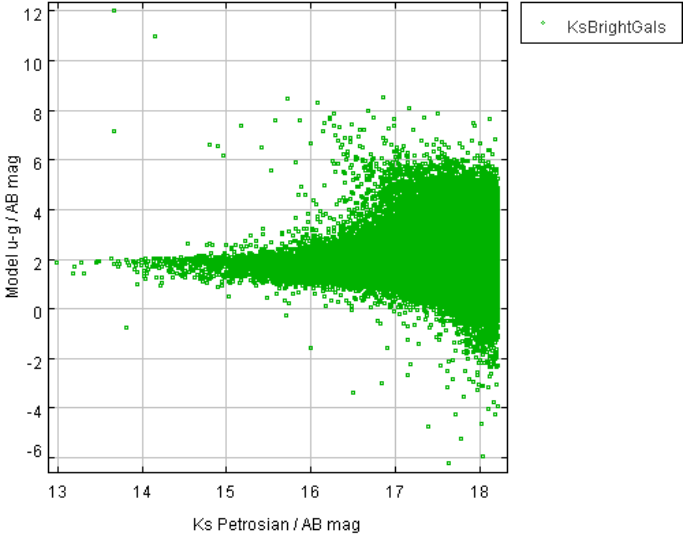}} 
\caption{Ks-band AB Petrosian magnitude versus u-g model mag colour for the 
 $Ks_{AB}<18.2$, $z<1$ sample selected in Fig~\ref{fig:KABvsRed}. At the bright 
end, the red-sequence is clear, but at fainter magnitudes, galaxies will be at 
higher redshift, so the observed colours are less meaningful.} 
\label{fig:KABvsCol} 
\end{figure}

\begin{figure} 
 \resizebox{\hsize}{!}{\includegraphics{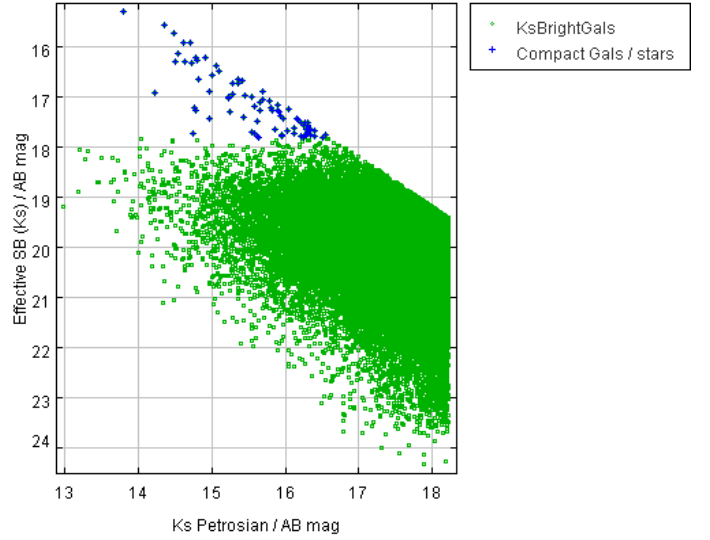}} 
 \caption{Ks-band AB Petrosian magnitude versus effective Ks surface brightness 
for the $Ks_{AB}<18.2$, $z<1$ sample selected in Fig~\ref{fig:KABvsRed}. The 
hard limit at the top-right hand side shows the size limit of $0.7\arcsec$. 
The main galaxy population seems to have a high surface brightness limit of 
$\mu_{Ks}=18.$ mag arcsec$^{-2}$, with a small group at higher surface 
 brightnesses (shown as blue crosses) - either compact galaxies or stars that 
have managed to avoid all the selection criteria. Galaxies have surface brightnesses as low as 
$\mu_{Ks}=24.0$ mag arcsec$^{-2}$, but fainter galaxies would need to be 
selected before the VIKING surface brightness became a limiting factor.} 
\label{fig:KABvsSB} 
\end{figure}

\newpage 
We plot the magnitude against the redshift for these galaxies,
Fig~\ref{fig:KABvsRed}, and find that there is an artificial selection at $z<1$.
The most likely explanation is that the photometric redshifts are limited to
this range, since the SDSS optical colours do not give reliable photometric
redshifts outside this range. By selecting a subsample at $K_{s}<18.2$ and
$z<1$, we have a more complete sample with reliable redshifts. We use this
sample to look at the colour-magnitude plot and the surface brightness magnitude
plot of galaxies, Figs~\ref{fig:KABvsCol} \& ~\ref{fig:KABvsSB}. The surface
brightness, colour, magnitude and redshift are all fundamental for classifying
galaxies.

\subsection{Extragalactic variables in VIDEO} 

In deep extragalactic surveys, such as VIDEO, with many epochs over several
months or years, it is possible to find a range of AGN, and very occasionally
supernovae. There are also a few foreground stars that show variability. We
select point-source variables in the VIDEO survey which show a range in
magnitudes greater than 0.1 mag in any filter. Since the filters in this survey are not
taken simultaneously, and AGN show sporadic variability, sometimes variations
may only be seen in one filter.

\begin{verbatim} 
SELECT s.sourceID,s.ra,s.dec,v.frameSetID, 
v.zMedianMag,v.zMagRms,v.znGoodObs,v.zSkewness, 
(v.zMaxMag-v.zMinMag) AS zRange,v.yMedianMag, 
v.yMagRms,v.ynGoodObs,v.ySkewness,  
(v.yMaxMag-v.yMinMag) AS yRange,v.jMedianMag, 
 v.jMagRms,v.jnGoodObs,v.jSkewness, 
 (v.jMaxMag-v.jMinMag) AS jRange,v.hMedianMag, 
v.hMagRms,v.hnGoodObs,v.hSkewness, 
(v.hMaxMag-v.hMinMag) AS hRange,v.ksMedianMag, 
v.ksMagRms,v.ksnGoodObs,v.ksSkewness, 
(v.ksMaxMag-v.ksMinMag) AS ksRange  
FROM videoVariability AS v,videoSource AS 
s /* join tables */ 
WHERE v.sourceID=s.sourceID AND 
/* point source variables */ 
 s.mergedClass IN (-1,-2) AND  
v.variableClass=1 AND  
 /* delta mag in > 0.1 in ANY filter, with  
at least 5 good obs in that filter */ 
(((zMaxMag-zMinMag)>0.1 AND zMinMag>0.  
AND znGoodObs>=5) OR ((yMaxMag-yMinMag)>0.1  
AND yMinMag>0. AND ynGoodObs>=5) OR  
((jMaxMag-jMinMag)>0.1 AND jMinMag>0. AND  
jnGoodObs>=5) OR ((hMaxMag-hMinMag)>0.1  
AND hMinMag>0. AND hnGoodObs>=5) OR  
((ksMaxMag-ksMinMag)>0.1 AND ksMinMag>0.  
 AND ksnGoodObs>=5))  
\end{verbatim} 

\begin{figure} 
\resizebox{\hsize}{!}{\includegraphics{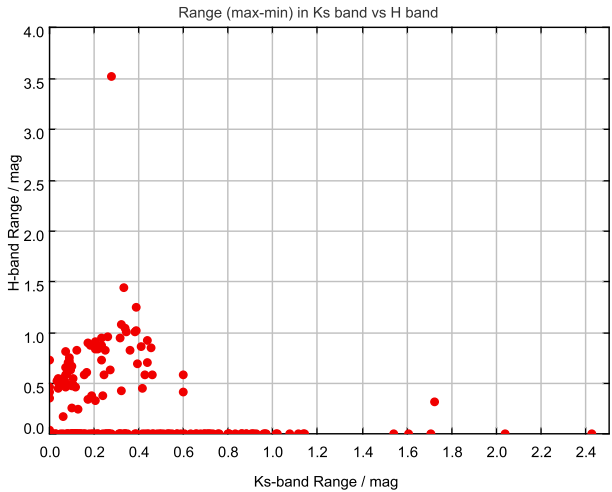}} 
 \caption{Ks-band magnitude range versus H-band magnitude range. While there 
 is a correlation between the two ranges, the H-band range seems greater than 
the Ks-band on average. There are some objects which have no discernible 
variation in Ks or H, but do so in one or more of the other filters.} 
\label{fig:RangeKsH} 
\end{figure}  

\begin{figure} 
\resizebox{\hsize}{!}{\includegraphics{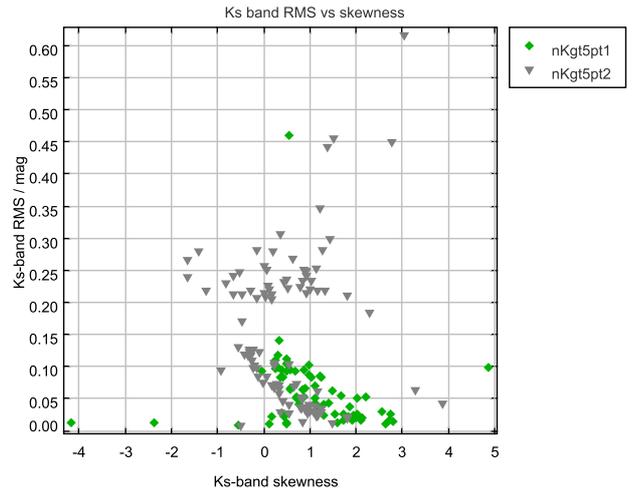}} 
 \caption{Ks-band Skewness versus RMS for objects with at least 5 good Ks-band 
observations. These have been split into the two pointings (pointing 1,
frameSetID=644245094401; pointing 2=644245094402), which show similarities, at
least for rms$<0.2$ mag. Most of the objects selected have positive skews in the Ks-band. The skew decreases as the RMS increases at
rms$<0.2$ mag, although the reason for this is not clear.}
\label{fig:RMSvsSkew} 
\end{figure}

We can plot some of the variability statistics, such as the range in the $K_{s}$
band against the range in the $H$ band, see Fig~\ref{fig:RangeKsH} or the
RMS against the skewness in the $K_{s}$ band, Fig~\ref{fig:RMSvsSkew}. These
types of plots help to classify different types of variable and to pick out odd
objects. 

We can then select one of these objects, e.g. the object in 
Fig~\ref{fig:RMSvsSkew} from pointing 1, with a $K_{s}$ band RMS $>0.4$ mag,
which is more than twice the RMS of any of the other objects and plot the light curve. To do this,
we do a second query, below:
	 
\begin{verbatim} 
SELECT  
/* Select time, filter, magnitude, magnitude  
error and flags */ 
d.mjd,d.filterID,d.aperMag3,d.aperMag3Err, 
 d.ppErrBits 
/* From BestMatch table to link all  
 observations of the same source and  
videoDetection for each observation */ 
 FROM videoDetection as d, 
videoSourceXDetectionBestMatch as b  
 /* First join tables */ 
 WHERE b.multiframeID=d.multiframeID AND  
b.extNum=d.extNum AND b.seqNum=d.seqNum 
 /* then select only detections and sourceID 
 equal to object in previous selection 
which has a Ks-band RMS>0.4 mag */ 
AND d.seqNum>0 AND b.sourceID IN (  
SELECT s.sourceID  
FROM videoVariability AS v,videoSource AS s  
 /* join tables */  
 WHERE v.sourceID=s.sourceID AND  
/* point source variables */  
s.mergedClass IN (-1,-2) AND  
v.variableClass=1 AND  
/* delta mag in > 0.1 in ANY filter, with  
at least 5 good obs in that filter */  
 (((zMaxMag-zMinMag)>0.1 AND zMinMag>0.  
AND znGoodObs>=5) OR ((yMaxMag-yMinMag)>0.1  
AND yMinMag>0. AND ynGoodObs>=5) OR  
 ((jMaxMag-jMinMag)>0.1 AND jMinMag>0. AND  
 jnGoodObs>=5) OR ((hMaxMag-hMinMag)>0.1 AND  
hMinMag>0. AND hnGoodObs>=5) OR  
 ((ksMaxMag-ksMinMag)>0.1 AND ksMinMag>0. AND 
ksnGoodObs>=5)) 
 /* Ks-band RMS >0.4 mag */ 
 AND ksMagRms>0.4 AND s.frameSetID=644245094401)  
 /* order by time */ 
 ORDER BY d.mjd 
 \end{verbatim} 
 
\begin{figure} 
\resizebox{\hsize}{!}{\includegraphics{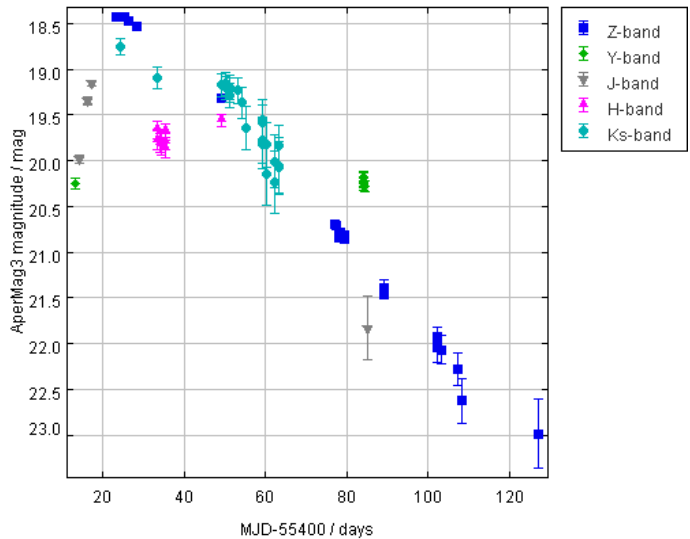}} 
\caption{Light-curve of variable selected in text. This variable brightens by 
$\sim2$ magnitudes in less than 10 days and then fades by almost 5 magnitudes 
over the next 100 days. This is the expected behaviour of a supernova Type 1a.} 
\label{fig:lcSNae} 
\end{figure}

\begin{figure}
\resizebox{\hsize}{!}{\includegraphics{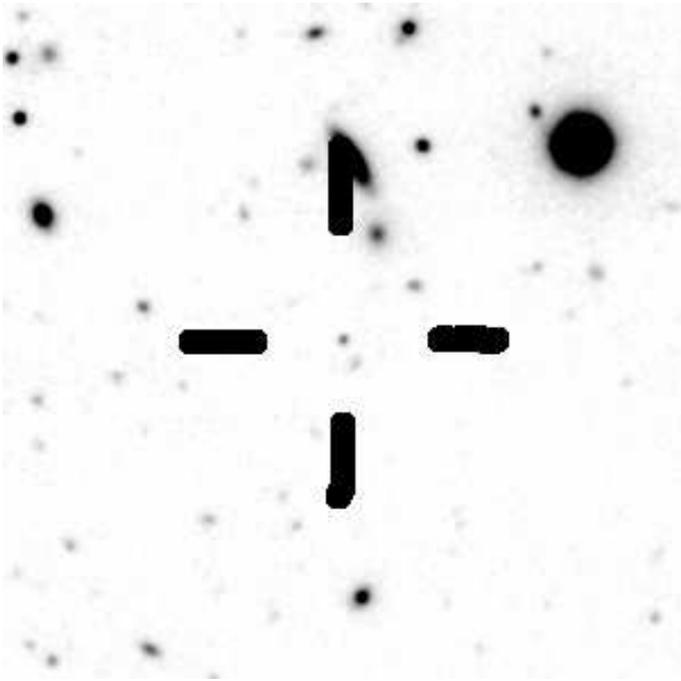}} 
\caption{Thumbnail of point-source variable which has light curve shown in 
Fig~\ref{fig:lcSNae}. Thumbnails can be shown by selecting the attributes 
{\bf ra}, {\bf dec} and {\bf frameSetID} from the {\texttt Source} table and 
clicking on the supplied link.} 
\label{fig:thumb}   
\end{figure}

We can use TOPCAT\footnote{http://www.star.bris.ac.uk/$\sim$mbt/topcat/} to plot
the light-curve, see Fig~\ref{fig:lcSNae}. The light-curve is very interesting, 
showing a short phase of brightening followed by a longer phase of fading, 
characteristic of an exploding star, probably a Type 1a SNa, with a maximum 
brightness of $Z=18.5$ mag. The position and time of this object match SN2010gy 
\citep{SN2010gy}. The discovery team found that it has a redshift, $z=0.06$, but
could not find a likely host galaxy. The thumbnail of this source from the deep 
$K_s$ band mosaic is shown in Fig~\ref{fig:thumb}.
  
\section{First Public Data Releases}
\label{sec:releases}

The first public releases of VISTA Public Survey data through the VSA are
intended to match the DR1/2 datasets published for each survey in the ESO 
SAF\footnote{http://www.eso.org/sci/observing/phase3/data$\_$releases.html}. 
Thus, they will cover the data up to the end of ESO semester P85 (i.e. up to
30th September 2010). Some of the surveys have released data from P86 as well,
but with the following additional constraints:

\begin{itemize}
  \item VMC: the data released will only be those fields where the whole set of
  epochs is complete, i.e. the 30~Doradus field (5h37m40s,-69$^{\circ}$
  22\arcmin18\arcsec) and the Gaia South Elliptical Pole field 
  (5h59m23s,-66$^{\circ}$20\arcmin28\arcsec).
  \item VIKING: the data will only be released in the following fields: GAMA09
  \citep[33 fields overlapping with the Galaxy and Mass Assembly 09h
  field; GAMA, ][]{GAMA}, CFHTLS-W1 (6 fields overlapping with the Canada
  France Hawaii Legacy Survey W1
  field\footnote{http://www.cfht.hawaii.edu/Science/CFHLS/} and 9 fields in
the Southern Galactic Pole region.
  \item VIDEO: the data will only be released in the pointings which VIDEO
  mosaics have been created in (XMM3 field, 2h26m18s, -4$^{\circ}$44\arcmin;
  ES1-North field, 0h37m49s, -43$^{\circ}$30\arcmin). 
\end{itemize}

We have cropped all the tables in the surveys to match the pointings specified
by the PIs. The excluded data will be released in future releases. 
Table~\ref{tab:dr1ProgSum} summarises the contents of the releases. Unlike UKIDSS
each survey will be released into a separate database. The sixth public survey
UltraVISTA is not using VDFS processing, except for the initial pawprint
pipeline calibration, so we are not releasing data from this survey.

\begin{table*}
\begin{center}
\caption{Summary of VISTA Public Survey DR1 VSA releases.}
\begin{tabular}{|l|r|c|c|c|} \hline
Survey & N(pointings) & Filters and depth$^a$ & N(epochs) & N(sources) \\ 
 &  & AB (mag) & (typical) & \\
\hline
\hline
VHS & 1260 & $Y\sim20.9$, $J\sim20.9$, $H\sim20.7$, $K_s\sim20.2$
& 1,1,1,1 & $1.2\times10^8$ \\ 
VVV & 350 & $Z\sim21.9$, $Y\sim20.9$, $J\sim20.8$, $H\sim20.2$,
$K_s\sim19.3$ & 1,1,1,1,6 & $5.0\times10^8$ \\ 
VMC & 2 & $Y\sim21.9$, $J\sim21.9$, $K_s\sim22.1$ & 4,4,12 &
$1.8\times10^6$ \\ 
VIKING & 48 & $Z\sim23.0$, $Y\sim21.9$, $J\sim21.8$,
$H\sim21.4$, $K_s\sim21.3$ & 1,1,2,1,1 & $5.3\times10^6$ \\  
VIDEO & 2 & $Z\sim25.6$, $Y\sim25.1$, $J\sim25.2$,
$H\sim24.8$, $K_s\sim24.5$ & 33,31,38,40,50 & $7.5\times10^5$ \\
\hline
\end{tabular}
\\ 
$^a$ for multi-epoch filters, this is calculated from the catalogue, but for a
single epoch, it is estimated from the exposure time.
\label{tab:dr1ProgSum}
\end{center}
\end{table*}

\section{Summary and future work}
\label{sec:summary}

The VSA was designed as the main access point to all VISTA science
data, allowing users to carefully select the data they need, rather than to 
bulk download all the data, a difficult and time consuming job in the era of 
billion row catalogues. As we have shown in the Illustrated Examples
(\S~\ref{sec:queries}), the VSA is designed to allow users to select on a wide
range of attributes and to work with external data, such as the Sloan Digital
Sky Survey, WISE, Glimpse, OGLE etc. The VSA is based on the WFCAM Science
Archive but has VISTA specific features and various improvements based on our
experience of WFCAM data and archive processing.

In the future we plan several enhancements. In the near future, we are working
on improvements to our interface, including a MyDB \citep{MyDB} style access,
where users can combine queries with Python scripts, to produce a powerful work 
environment. We are also improving our plotting tools to more easily show very 
large datasets with a combination of density maps where the number of points is
huge, and individual points where the density drops below a threshold.


\section{Acknowledgements}  \label{acknow}
 
Many people have contributed to the development of the VISTA Science
Archive. We would like to thank the PIs and members of the VISTA Public Survey
teams who have made requests, spotted mistakes, and checked the early releases
carefully. We would also like to thank users of the WFCAM Science Archive.
We thank Steve Warren for providing Figure~\ref{fig:WFCAMfp}. We would like to
thank the anonymous referee for his/her careful reading of the text and
useful suggestions that have improved this paper. We would also like to thank
Pratika Dayal for pointing out the VISTA/VDFS jargon that we have got so 
used to using.

We thank Microsoft also for
providing software via the Microsoft Development Network Academic
Alliance, and eGenix for providing a free, non-commercial licence
for Python middleware. 

Financial resources for VISTA/WFCAM Science Archive development and operations were
provided by the UK Science and Technology Facilities Council.

\

\bibliographystyle{aa}
\bibliography{vsapaper}

\appendix

\section{Image scaling}
\label{app:scale}

For single epoch OB stacks,

\begin{equation}
bscale=\frac{1}{NDIT\,\sqrt{NJITTER}},
\end{equation}

\noindent and for tiles,
\begin{equation}
bscale=\frac{1}{NDIT\,\sqrt{2\,NJITTER}},
\end{equation}

\noindent where $NDIT$ is the number of readouts during the integration of a raw
image and $NJITTER$ is the number of jitter positions in the jitter pattern to
create a single pawprint stack. For deep stacks, we would scale by the number of epochs,
in the same way as the number of jitters, but since $NDIT$ and $NJITTER$
can vary from stack to stack in the same programme, pointing and filter, we
calculate $bscale$ for deep stacks as:

\begin{equation}
bscale^{\rm deep}=\frac{1}{\sqrt{\sum_i\,\frac{1}{bscale_i^2}}}
\end{equation}

For deep tiles, to take account of different integration times in each offset
and deprecated detectors in some OB stacks, we compute the $bscale$ values for
each detector as above in each overlap and average over all overlaps, just as we
calculate the total exposure time for a tile. 

\section{Half-light radii}
\label{app:hlr}

The sizes of extended sources are difficult to measure for various reasons:

\begin{itemize}
  \item The outer parts of a galaxy eventually merge into the sky background, so
  it is difficult to know how much of the galaxy is lost in the sky. For
  intrinsically low surface brightness galaxies or high redshift galaxies, the
  majority of the light may well be lost in the background and any measurement
  is a significant underestimate \citep[e.g.]{Dis76,Crs01}
  \item The profile is not always smooth or axisymmetrical, e.g. grand design
  spirals, irregular galaxies, interacting galaxies.
  \item Nearby objects can make it difficult to get accurate measurements of the
  total luminosity and extent of a galaxy. The measurement of the background
  level is sometimes incorrect and this can affect a curve-of-growth
  measurement. 
  \item Galaxies have various inclinations to the line of sight. Different
  researchers may want to use measurements that correct for inclination or do
  not.
  \item The light of galaxies and all objects is convolved with a point-spread
  function that will particularly effect small objects.
  \item The method must be robust enough and quick enough to be applied to the
  VISTA detection tables. We exclude the VVV and VMC since these are in
  extremely dense regions where contamination from nearby objects is almost
  guaranteed and the vast majority of sources are stellar, which are
  point-sources. Even so, the catalogues with size measurements will contain
  $>10^8$ sources, and maybe $10^9$ sources. 
\end{itemize}

Our measurements of the size of galaxies try to take into account all the above
effects as much as possible. We define our basic size measurement as the radius
containing half of the flux of the galaxy -- the half-light-radius -- a
measurement that has been used extensively before \citep{Kor77,Crs01,Blt01}. The
main difficulty with this measurement is measuring the total flux.  
 
To take into account the missing light from the outer parts of the galaxy, we
use the Petrosian flux \citep{Blt01,Grh05a}, which is generally insensitive to
the effects of surface brightness, i.e. if you keep the galaxy profile the same
(the relative flux as a function of radius) but reduce or increase the average
surface-brightness, closer to or further from the sky noise value, then the
Petrosian flux measurement will return the same flux each time. This breaks 
down eventually: if you reduce the surface brightness enough the galaxy won't 
even be detectable against the sky, and close to this limit the total flux and
size will become difficult to measure with any accuracy.

The Petrosian flux however gives different results for different profiles, which
is an issue. \cite{Blt01} showed that while only 0.7\% of the flux of an
exponential disk galaxy was typically missed by the Petrosian, 22\% of the flux
of a de Vaucouleurs' profile elliptical galaxy was missed, and 5\% of the flux
of a PSF dominated profile was missed, although \cite{Grh05b} shows that there
are slightly different results for a standard Petrosian definition compared to
the SDSS Petrosian that Blanton used. Small galaxies, close to the seeing
limit will be dominated by the point-spread function. Galaxies close to the 
surface-brightness limit of the survey could be missing much more of the light.
To try to take into account the missing light, we assume that all galaxies are 
missing 10\%. This will be an overestimate in some cases and an underestimate in
others, but to try and calculate a correction for each galaxy would require an 
iterative procedure which would take much longer, and in any case, it would be 
better to fit profiles for all objects \citep[e.g.][]{GALFIT}. The light
profiles of galaxies are often well fit by S\'ersic profiles \citep{Grh05a}, a more
general function, that includes exponential ($\beta=1$), de Vaucouleur
($\beta=4$) and Gaussian ($\beta=0.5$),

\begin{equation}
I(r)=I_{r_{hl}}\exp{\left\{-k\left[\left(\frac{r}{r_{hl}}\right)^{1/\beta}-1\,\right]\,\right\}}
\end{equation}

\noindent where $r_{hl}$ is the half-light radius, $\beta$ is the S\'ersic
index. To save time, we use the existing catalogue products to calculate the
half-light radii, rather than going back to the images \citep[e.g.][]{Lsk03}. 
The half-light radii are calculated using the existing circular
aperture radii measurements of the flux, which give a curve-of-growth. We use
the 13 aperture fluxes measured by the VDFS extractor, at radii of 
$0.5, 0.5\sqrt{2}, 1, \sqrt{2}, 2, 2\sqrt{2}, 4, 5, 6, 7, 8, 10, 12$
arcseconds. To calculate the half-light radius, we first find the aperture flux closest to
half the total-flux and then use the five apertures centred on this (2 before, 
2 after and the aperture in question). Using these 5 aperture fluxes, we fit a 
quadratic, which removes any small bumps in the curve, using the singular value
decomposition method \citep{SVD}. We find the
root of the quadratic that gives the half-light radius ($r_{hl}^c$, {\bf
hlCircRadAs}) . We use the covariance matrix to calculate the error in the half-light radius
($\sigma_{r_{hl}^c}$, {\bf hlCircRadErrAs}), adding in another half-pixel in
quadrature, to take into account the granularity of the data.

\begin{figure}
\resizebox{\hsize}{!}{\includegraphics{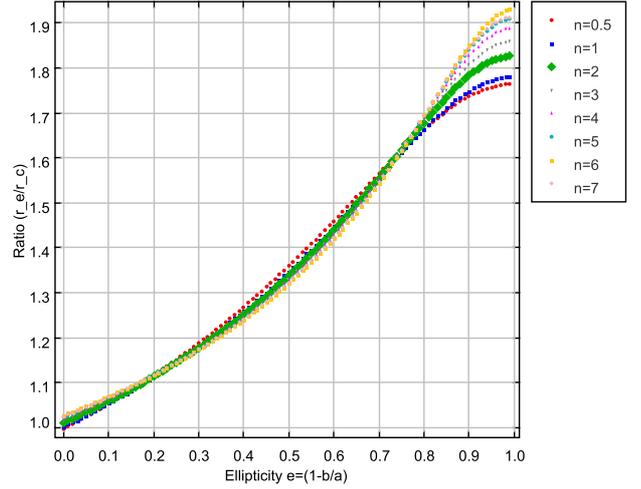}}
  \caption{Figure of the ellipticity versus ratio of half-light semi-major axis
  to half-light radius for different S\'ersic profiles. The lines are the best
  fit Moffat profiles in each case.}
\label{fig:rellrcirc}
\end{figure}

The 13 aperture fluxes are all circular apertures, so the half-light radius
calculated assumes a circular symmetry. However, most galaxies are elliptical 
in shape, either being triaxial spheroidal systems or inclined disks or a 
combination of the two. A geometric mean size is usually considered a
more useful measurement for triaxial elliptical galaxies \citep[e.g.
see][]{BinMer}, and a semi-major axis size \citep{Drv05}, which recovers the 
radius of the disk whatever the inclination, is more useful for disk galaxies.
Figure~\ref{fig:rellrcirc} shows
the ratio of half-light semi-major axis to half-light circular radius as a function of ellipticity for a range of S\'ersic
profiles ($\beta=0.5$ to $\beta=7$): profiles which are a good fit to the vast
majority of virialised galaxies. As can be seen, the variation between these profiles is
around $1-2\%$ for all ellipticities $<0.9$, but rises to $~10\%$ at maximum
ellipticity. These curves are well fit by Moffat functions,

\begin{equation}
f(\epsilon)=\frac{a}{(1+(\frac{1-\epsilon}{b})^2)^c}
\end{equation}

\noindent where $a$, $b$ \& $c$ are found from fitting the data for each
profile.

Thus, we can convert our circular half-light radii $r_{hl}^c$ to a semi-major
axis size $r_{hl}^{smj}$ by using the $\beta=2$ function
($a=1.8243,b=0.30914,c=0.24304$). We choose $\beta=2$, since most galaxies will
be ellipticals ($\beta=1$), de Vaucouleurs ($\beta=4$), or dominated by the PSF
($\beta=0.5$), and $\beta=2$ is nicely in the middle, but as
Fig~\ref{fig:rellrcirc} shows, there is very little difference. 

The conversions to the half-light semi-minor axis, $r_{hl}^{smn}$, and
half-light geometric mean, $r_{hl}^{geo}$, are easily computed from the geometry
of an ellipse:

\begin{equation}
r_{hl}^{smn}=(1-\epsilon)r_{hl}^{smj}
\end{equation}

\begin{equation} 
r_{hl}^{geo}=\sqrt{r_{hl}^{smn}r_{hl}^{smj}}
\end{equation} 

Finally, we take into account the effects of seeing. We use the method of
\cite{Drv05} to subtract the measured seeing, assuming a Gaussian PSF, 

\begin{equation}
r_{hl}^{smj,see}=\sqrt{{r_{hl}^{smj}}^2-c_{see}\Gamma^2},
\end{equation}

\noindent where $\Gamma$ is the full-width half maximum of stars in the image
and $c_{see}$ is a constant, 0.5 for a Gaussian PSF. By experiment we found values of
$c_{see}\sim0.45$, but with quite a large uncertainty, so we took the
theoretical value $c_{see}=0.5$.

\label{lastpage}

\end{document}